\documentclass[twocolumn]{aastex702}

\usepackage{comment}
\usepackage{xcolor}

\begin{document}

\title{Population II Post-AGB Stars. I. Methodology, Initial Results, and the First Luminosity Function} 

\author[orcid=0000-0002-0424-8519, sname='Feeser']{S. Jean Feeser}
\affiliation{Department of Astronomy and Astrophysics, The Pennsylvania State University, University Park, PA, 16802, USA}
\email[show]{sjfeeser@psu.edu}  

\author[orcid=0000-0002-1328-0211, sname='Ciardullo']{Robin Ciardullo}
\altaffiliation{Institute for Gravitation and the Cosmos, The Pennsylvania State University, University Park, PA, 16802, USA}
\affiliation{Department of Astronomy and Astrophysics, The Pennsylvania State University, University Park, PA, 16802, USA}
\email{rbc3@psu.edu}

\author[orcid=0000-0003-1817-3009, sname='Siegel']{Michael H. Siegel}
\affiliation{Department of Astronomy and Astrophysics, The Pennsylvania State University, University Park, PA, 16802, USA}
\email{mhs18@psu.edu}

\author{C. J. Hapich}
\affiliation{Department of Astronomy and Astrophysics, The Pennsylvania State University, University Park, PA, 16802, USA}
\email{Charles.Hapich01@student.csulb.edu}

\author[orcid=0009-0008-0451-4435, sname='Chaturvedi']{Akshat S. Chaturvedi}
\altaffiliation{Center for High Angular Resolution Astronomy and Department of Physics and Astronomy, Georgia State University, 25 Park Place NE, Suite 605, Atlanta, GA 30303, USA}
\affiliation{Department of Astronomy and Astrophysics, The Pennsylvania State University, University Park, PA, 16802, USA}
\email{achaturvedi3@gsu.edu}

\defcitealias{Davis2022}{D22}
\defcitealias{Bertolami2016}{MB16}
\defcitealias{Moehler2019}{M19}
\defcitealias{Cardelli1989}{CCM89}
\defcitealias{Vasiliev2021}{VB21}
\defcitealias{Zinn1972}{ZNG72}

\begin{abstract}
Post-asymptotic giant branch (PAGB) stars share many advantages with the tip of the red giant branch (TRGB) for Population II distance measurements, including minimal internal reddening, easy identification, a solid theoretical foundation, and an expected weak metallicity dependence. At visual wavelengths PAGB stars are the brightest members in old populations, appearing up to $\sim2$ mag brighter than the TRGB and $\sim4$ mag brighter than RR~Lyrae stars. Stellar evolution theory predicts that these objects occupy a narrow luminosity strip across the top of the Hertzsprung-Russell diagram, making them excellent standard candles. Observations in Galactic globular clusters (GCs) support this prediction, but a comprehensive survey is required to define their luminosity function and test for systematic errors.

In this work, we construct a multi-wavelength catalog, spanning the ultraviolet (UV) to infrared (IR), for PAGB stars in 39 Galactic GCs. We validate cluster membership with \textit{Gaia} astrometry, identify 21 objects across 16 GCs using ground-based \textit{uBVI} and space-based UV photometry, and perform Bayesian spectral energy distribution (SED) fitting with stellar atmosphere models to estimate their effective temperatures, luminosities, and radii. We find that PAGB stars occupy a narrow luminosity range centered at $\log(L_{\mathrm{bol}}/L_{\odot})=3.25\pm0.03$, with dispersion $\sigma=0.12$ dex, supporting their use as reliable Population II distance indicators. Comparisons with theoretical tracks show excellent agreement with evolutionary models. We confirm that PAGB stars occur preferentially in metal-poor clusters with blue horizontal branches. Our results demonstrate that PAGB stars provide a powerful probe of late stellar evolution and a promising Population II standard candle.
\end{abstract}


\section{Introduction}\label{sec:intro}
Globular clusters (GCs) are compact, gravitationally bound systems of thousands to millions of stars and are among the oldest stellar populations in the Milky Way. While GCs are known to host multiple stellar populations with internal variations of light element abundances \citep{Kraft1994, Gratton2004, Bastian2018}, their global properties (such as ages, metallicities, and distances) are well-constrained and exhibit very narrow distributions, making them ideal `laboratories' for advancing our knowledge of stellar structure and evolution \citep{Harris1996, Harris2010}. They also play a central role in the cosmic distance scale, hosting and providing calibration for standard candles such as RR Lyrae and the tip of the red giant branch (TRGB) \citep[e.g.,][]{Braga2018, Muraveva2018, Freedman2021}.

In color-magnitude diagrams (CMDs), GCs display well-defined evolutionary sequences characteristic of old, low-mass stellar populations, including a prominent red giant branch (RGB), horizontal branch (HB), and asymptotic giant branch (AGB\null).  Of special importance are the few luminous members that populate the brightest regimes of ultraviolet (UV) and optical CMDs, many of which are stars in their final stages of evolution \citep{Zinn1972, deBoer1985}. Among these, post-asymptotic giant branch (PAGB) stars have emerged as promising Population II standard candles and valuable probes of post-HB stellar evolution. However, systematic studies of PAGB stars have been limited by uncertain cluster membership and incomplete photometric coverage. With the advent of precise astrometry from \textit{Gaia} and the availability of wide-field UV imaging from space-based missions, such as the Neil Gehrels \textit{Swift} Observatory \citep{Roming2005} and The Galaxy Evolution Explorer \citep[GALEX;][]{Morrissey2007}, it is now possible to overcome these limitations and complete a reliable census of PAGB stars across Galactic GCs.

\citet{Bond1997} first proposed that PAGB stars in GCs satisfy many of the criteria for an excellent Population II standard candle: they are intrinsically bright, are minimally affected by reddening, show weak metallicity dependence, and should exhibit a very small scatter in bolometric luminosity.  In their search for intermediate-temperature above-horizontal-branch (AHB) stars, \citet{Davis2022} (hereafter D22)   identified 13 PAGB stars brighter than $M_V \simeq -2.75$ in 100 Milky Way GCs.  Among these were five non-variable yellow PAGB (yPAGB) stars with a mean absolute magnitude $M_V = -3.31 \pm 0.05$ and a standard deviation of just 0.11 mag.  \citet{Ciardullo2022} then confirmed this value by doubling the sample size via the inclusion of the intensity-weighted mean magnitudes of five RV Tauri variables. They found a mean bolometric magnitude of $\langle M_{\mathrm{bol}} \rangle = -3.38 \pm 0.03$ for the combined dataset, with a dispersion of 0.10 mag.  These results suggested that yPAGB stars could deliver distance measurements that are accurate to $\sim 2\%$ out to $\sim 10$ Mpc.

D22's analysis was limited by the small number of intermediate-temperature ($T_{\mathrm{eff}} \sim 5{,}000$ to $10{,}000$ K) PAGB stars in GCs.  However, many more of these extremely luminous objects are present at higher temperatures, where their UV colors clearly separate them from most other GC stars.   

Examining the full temperature range of PAGB stars can also provide critical tests of late-stage stellar evolution models. Recent predictions suggest that PAGB evolution may be faster and more luminous than previously assumed, with \citet{Bertolami2016} (hereafter MB16) estimating timescales that are three to ten times shorter and $\sim 0.1\,\text{--}\,0.3$ dex brighter than older models. Additionally, theoretical tracks by \citet{Moehler2019} (hereafter M19) predict that the frequency of PAGB stars are strongly correlated with a cluster's HB morphology. Although the observed numbers of PAGB and post-early AGB (PEAGB) stars appear consistent with these models, meaningful comparisons have been limited by small sample sizes within a narrow temperature range. 

Finally, PAGB stars are a significant source of UV radiation in older stellar systems and partially contribute to the observed UV upturn of elliptical galaxies \citep{OConnell1999, Yi2008, Siegel2024}. A better empirical understanding of their luminosity function and evolutionary lifetimes is needed for accurate population synthesis modeling, interpreting the integrated light of unresolved stellar populations, and characterizing the UV emission observed in early-type galaxies.

This series aims to perform a complete census of PAGB stars in Galactic GCs, extending previous efforts that identified the intermediate-temperature population.  Our ultimate goal is to define the objects' luminosity and temperature distributions, examine their connection with HB morphology, test models of PAGB stellar evolution, and refine their utility as extragalactic distance indicators.  In this, the first paper of our series, we develop the methodology required to achieve this goal and apply it to a sample of 39 GCs spanning a range of HB morphologies, metallicities, and distances. Section~\ref{sec:data} describes the multi-wavelength photometric catalog and astrometric criteria used to identify cluster members. Section~\ref{sec:PAGBcensus} outlines the PAGB candidate selection and identifies potential PAGB stars. Section~\ref{sec:SED} describes the spectral energy distribution (SED) fitting procedure, interprets the results, and discusses limitations of the analysis. Section~\ref{sec:Lfunction} evaluates these results using theoretical evolutionary models and examines cluster-to-cluster variations. Section~\ref{sec:discussion} compares the inferred stellar properties with previously published values and reviews PAGB discoveries in the literature. Finally, Section~\ref{sec:future} summarize this work, its implications, and outlines directions for future investigations.

\section{Observations and Membership Selection}\label{sec:data}
In this initial study, we selected 39 GCs spanning a broad range of metallicities, luminosities, and HB morphology. The target list is given in Table~\ref{tab:GCParameters}. Almost all clusters were imaged from the UV through the IR using exclusively wide-field instruments. These observations were not extremely deep, but designed to provide as complete a census as possible, with accurate photometry of evolved AHB stars from the cluster center to approximately their tidal radius when feasible. This objective was achieved for most clusters. However, due to high stellar densities in the cluster cores, crowding may reduce AHB completeness in this region.

\begin{longrotatetable}
\begin{deluxetable*}{lccccccccccl}
\tabletypesize{\scriptsize}
\tablewidth{0pt}
\tablecaption{Globular Cluster Parameters \label{tab:GCParameters}}
\tablehead{
\colhead{Cluster ID} & \colhead{Other Name} & \colhead{RA} & \colhead{Dec} & \colhead{$\mu_{\alpha}$} & \colhead{$\mu_{\delta}$} & \colhead{Distance} & \colhead{$E(B-V)$} & \colhead{$dE(B-V)^a$} & \colhead{[Fe/H]} & \colhead{HBR$^b$} & \colhead{References} \\
 &  & (J2000) & (J2000) & (mas yr$^{-1}$) & (mas yr$^{-1}$) & (kpc) & (mag) & (mag) & (dex) &  &  \\
}
\startdata
E 3 & ESO 37-1 & 09 20 57.12 & $-$77 16 55.2 & $-$2.73 & \phantom{$-$}7.08 & 8.100 & 0.30 & 0.016 & $-$0.83 & --- & H10; VB21; P24; L94 \\
IC 4499 & --- & 15 00 18.48 & $-$82 12 50.4 & \phantom{$-$}0.47 & $-$0.49 & 18.880 & 0.23 & 0.035 & $-$1.53 & \phantom{$-$}0.11 & H10; VB21; BV21; P24; L94 \\
NGC 288 & --- & 00 52 45.12 & $-$26 34 58.8 & \phantom{$-$}4.16 & $-$5.71 & 8.983 & 0.03 & 0.015 & $-$1.32 & \phantom{$-$}0.98 & H10; VB21; BV21; P24; L94 \\
NGC 362 & --- & 01 03 14.16 & $-$70 50 56.4 & \phantom{$-$}6.69 & $-$2.54 & 8.770 & 0.05 & --- & $-$1.26 & $-$0.87 & H10; VB21; BV21; L94 \\
NGC 1261 & --- & 03 12 16.32 & $-$55 12 57.6 & \phantom{$-$}1.60 & $-$2.06 & 16.353 & 0.01 & 0.006 & $-$1.27 & $-$0.71 & H10; VB21; BV21; P24; L94 \\
NGC 1851 & --- & 05 14 06.72 & $-$40 02 49.2 & \phantom{$-$}2.15 & $-$0.65 & 12.017 & 0.02 & 0.026 & $-$1.18 & $-$0.36 & H10; VB21; BV21; P24; L94 \\
NGC 1904 & M 79 & 05 24 10.56 & $-$24 31 26.4 & \phantom{$-$}2.47 & $-$1.59 & 13.080 & 0.01 & 0.007 & $-$1.60 & \phantom{$-$}0.89 & H10; VB21; BV21; P24; L94 \\
NGC 2298 & --- & 06 48 59.52 & $-$36 00 18.0 & \phantom{$-$}3.32 & $-$2.17 & 9.890 & 0.14 & 0.076 & $-$1.92 & \phantom{$-$}0.93 & H10; VB21; BV21; P24; L94 \\
NGC 2419 & --- & 07 38 08.40 & +38 52 55.2 & \phantom{$-$}0.01 & $-$0.52 & 88.471 & 0.08 & --- & $-$2.15 & \phantom{$-$}0.86 & H10; VB21; BV21; L94 \\
NGC 2808 & --- & 09 12 03.12 & $-$64 51 46.8 & \phantom{$-$}0.99 & \phantom{$-$}0.27 & 10.139 & 0.22 & 0.047 & $-$1.14 & $-$0.49 & H10; VB21; BV21; P24; L94 \\
NGC 3201 & --- & 10 17 36.72 & $-$46 24 43.2 & \phantom{$-$}8.35 & $-$1.96 & 4.749 & 0.24 & 0.189 & $-$1.59 & \phantom{$-$}0.08 & H10; VB21; BV21; P24; L94 \\
NGC 4590 & M 68 & 12 39 28.08 & $-$26 44 38.4 & $-$2.74 & \phantom{$-$}1.78 & 10.409 & 0.05 & 0.022 & $-$2.23 & \phantom{$-$}0.17 & H10; VB21; BV21; P24; L94 \\
NGC 4833 & --- & 12 59 33.84 & $-$70 52 37.2 & $-$8.38 & $-$0.96 & 6.519 & 0.32 & 0.132 & $-$1.85 & \phantom{$-$}0.93 & H10; VB21; BV21; P24; L94 \\
NGC 5024 & M 53 & 13 12 55.20 & +18 10 04.8 & $-$0.13 & $-$1.33 & 18.518 & 0.02 & 0.024 & $-$2.10 & \phantom{$-$}0.81 & H10; VB21; BV21; P24; L94 \\
NGC 5053 & --- & 13 16 27.12 & +17 42 00.0 & $-$0.33 & $-$1.21 & 17.515 & 0.01 & 0.012 & $-$2.27 & \phantom{$-$}0.52 & H10; VB21; BV21; P24; L94 \\
NGC 5272 & M 3 & 13 42 11.52 & +28 22 37.2 & $-$0.15 & $-$2.67 & 10.167 & 0.01 & 0.015 & $-$1.50 & \phantom{$-$}0.08 & H10; VB21; BV21; P24; L94 \\
NGC 5286 & --- & 13 46 26.88 & $-$51 22 26.4 & \phantom{$-$}0.20 & $-$0.15 & 11.071 & 0.24 & 0.079 & $-$1.69 & \phantom{$-$}0.80 & H10; VB21; BV21; P24; L94 \\
NGC 5466 & --- & 14 05 27.36 & +28 32 02.4 & $-$5.34 & $-$0.82 & 16.106 & 0.00 & --- & $-$1.98 & \phantom{$-$}0.58 & H10; VB21; BV21; L94 \\
NGC 5634 & --- & 14 29 37.20 & $-$05 58 33.6 & $-$1.69 & $-$1.48 & 25.990 & 0.05 & 0.016 & $-$1.88 & --- & H10; VB21; BV21; P24; L94 \\
NGC 5824 & --- & 15 03 58.56 & $-$33 04 04.8 & $-$1.19 & $-$2.23 & 31.754 & 0.13 & 0.015 & $-$1.91 & \phantom{$-$}0.79 & H10; VB21; BV21; P24; L94 \\
NGC 5897 & --- & 15 17 24.48 & $-$21 00 36.0 & $-$5.42 & $-$3.39 & 12.607 & 0.09 & --- & $-$1.90 & \phantom{$-$}0.86 & H10; VB21; BV21; L94 \\
NGC 5904 & M 5 & 15 18 33.12 & +02 04 51.6 & \phantom{$-$}4.09 & $-$9.87 & 7.471 & 0.03 & 0.017 & $-$1.29 & \phantom{$-$}0.31 & H10; VB21; BV21; P24; L94 \\
NGC 5927 & --- & 15 28 00.72 & $-$50 40 22.8 & $-$5.06 & $-$3.22 & 8.215 & 0.45 & 0.248 & $-$0.49 & $-$1.00 & H10; VB21; BV21; P24; L94 \\
NGC 5986 & --- & 15 46 02.88 & $-$37 47 09.6 & $-$4.19 & $-$4.57 & 10.549 & 0.28 & 0.103 & $-$1.59 & \phantom{$-$}0.97 & H10; VB21; BV21; P24; L94 \\
NGC 6093 & M 80 & 16 17 02.40 & $-$22 58 33.6 & $-$2.93 & $-$5.58 & 10.342 & 0.18 & --- & $-$1.75 & \phantom{$-$}0.93 & H10; VB21; BV21; L94 \\
NGC 6101 & --- & 16 25 48.00 & $-$72 12 07.2 & \phantom{$-$}1.76 & $-$0.26 & 14.454 & 0.05 & 0.019 & $-$1.98 & \phantom{$-$}0.84 & H10; VB21; BV21; P24; L94 \\
NGC 6171 & M 107 & 16 32 31.92 & $-$13 03 14.4 & $-$1.94 & $-$5.98 & 5.629 & 0.33 & --- & $-$1.02 & $-$0.73 & H10; VB21; BV21; L94 \\
NGC 6205 & M 13 & 16 41 41.28 & +36 27 36.0 & $-$3.15 & $-$2.57 & 7.427 & 0.02 & 0.017 & $-$1.53 & \phantom{$-$}0.97 & H10; VB21; BV21; P24; L94 \\
NGC 6218 & M 12 & 16 47 14.16 & $-$01 56 56.4 & $-$0.19 & $-$6.80 & 5.136 & 0.19 & 0.044 & $-$1.37 & \phantom{$-$}0.97 & H10; VB21; BV21; P24; L94 \\
NGC 6273 & M 19 & 17 02 37.68 & $-$26 16 04.8 & $-$3.25 & \phantom{$-$}1.66 & 8.527 & 0.38 & --- & $-$1.74 & \phantom{$-$}0.86 & H10; VB21; BV21; L94 \\
NGC 6352 & --- & 17 25 29.04 & $-$48 25 19.2 & $-$2.16 & $-$4.45 & 5.534 & 0.22 & --- & $-$0.64 & $-$1.00 & H10; VB21; BV21; L94 \\
NGC 6356 & --- & 17 23 35.04 & $-$17 48 46.8 & $-$3.75 & $-$3.39 & 15.900 & 0.28 & --- & $-$0.40 & $-$1.00 & H10; VB21; BV21; L94 \\
NGC 6402 & M 14 & 17 37 36.00 & $-$03 14 45.6 & $-$3.59 & $-$5.06 & 9.204 & 0.60 & --- & $-$1.28 & \phantom{$-$}0.65 & H10; VB21; BV21; L94 \\
NGC 6496 & --- & 17 59 03.60 & $-$44 15 57.6 & $-$3.06 & $-$9.27 & 9.647 & 0.15 & --- & $-$0.46 & $-$1.00 & H10; VB21; BV21; L94 \\
NGC 6637 & M 69 & 18 31 23.04 & $-$32 20 52.8 & $-$5.03 & $-$5.83 & 8.880 & 0.18 & --- & $-$0.64 & $-$1.00 & H10; VB21; BV21; L94 \\
NGC 6652 & --- & 18 35 45.60 & $-$32 59 27.6 & $-$5.48 & $-$4.27 & 9.467 & 0.09 & --- & $-$0.81 & $-$1.00 & H10; VB21; BV21; L94 \\
NGC 6712 & --- & 18 53 04.32 & $-$08 42 21.6 & \phantom{$-$}3.36 & $-$4.44 & 7.355 & 0.45 & 0.266 & $-$1.02 & $-$0.62 & H10; VB21; BV21; P24; L94 \\
NGC 6779 & M 56 & 19 16 35.52 & +30 10 58.8 & $-$2.02 & \phantom{$-$}1.62 & 10.404 & 0.26 & --- & $-$1.98 & \phantom{$-$}0.98 & H10; VB21; BV21; L94 \\
NGC 7089 & M 2 & 21 33 27.12 & $-$00 49 22.8 & \phantom{$-$}3.44 & $-$2.16 & 11.647 & 0.06 & 0.017 & $-$1.65 & \phantom{$-$}0.96 & H10; VB21; BV21; P24; L94 \\
\enddata
\tablecomments{
Units of right ascension are hours, minutes, seconds; units of declination are degrees, arcminutes, arcseconds. \\
$^a$ Cluster differential reddening, $dE(B-V)$, adopted from \citet{Pancino2024}. \\
$^b$ Cluster Horizontal Branch Ratio (HBR), defined by Equation~\ref{eq:HBR} and described in Section~\ref{sec:postHBevolution}. \\
References are abbreviated as follows: H10, \citet{Harris2010} for cluster parameters (reddening and metallicity); VB21, \citet{Vasiliev2021} for proper motions and cluster parameters; BV21, \citet{Baumgardt2021} for updated distances; P24, \citet{Pancino2024} for $dE(B-V)$; L94, \cite{Lee1994}, BM00, \cite{Borkova2000}, or C09, \cite{Catelan2009} for HBR. \\
}
\end{deluxetable*}
\end{longrotatetable}

\subsection{Photometric Data}\label{sec:photo}
Among the AHB population, intermediate-temperature stars have SEDs characterized by an extremely large Balmer decrement due to their low surface gravities. The \textit{uBVI} photometric system was established by \citet{Bond2005} as an effective tool for measuring this feature.  In the system, the  \citet{ThuanGunn1976} $u$ filter, which lies entirely blueward of the Balmer jump ($\lambda_{\mathrm{eff}} = 3530$ \AA), is compared to the Johnson-Kron-Cousins (JKC) $B$-filter to define the amplitude of the break, while the JKC $V$ and $I$ filters provide an estimate of the stellar temperature. \citet{Bond2005} explains, in detail, the scientific motivation for this photometric system, including its filter properties and corrections, and calibration was provided by \citet{Siegel2005}.

To build a comprehensive multi-wavelength catalog of bright (above the HB) globular cluster stars, we began with the data of \citetalias{Davis2022}, which provides \textit{uBVI} photometry for 100 Galactic GCs and nine Population II GCs in the Magellanic Clouds. We then cross-matched the coordinates of objects with data from the following sources, when available:
\begin{enumerate}
    \item The Galaxy Evolution Explorer \citep[\textit{GALEX};][]{Morrissey2007}.
    This satellite provides far-UV ($\lambda_{\mathrm{eff}} = 1516$~\AA) and near-UV ($\lambda_{\mathrm{eff}} = 2267$~\AA) photometry for 44 GCs \citep{Schiavon2012}.
   
    \item The Ultra-Violet/Optical Telescope \citep[UVOT;][]{Roming2005} on the Neil Gehrels \textit{Swift} Observatory. We include observations taken through the \textit{uvw2}, \textit{uvm2}, and \textit{uvw1} UV filters ($\lambda_{\mathrm{eff}} = 2030$ \AA, $2231$ \AA, and $2634$ \AA, respectively) and optical $u$ filter ($\lambda_{\mathrm{eff}} = 3501$ \AA), obtained as part of the ongoing survey program described in \citet{Siegel2014}. \textit{Swift} photometry was measured using \texttt{DAOPHOT} using techniques described in \citet{Siegel2014}. The full catalog will be presented in a future contribution.
    
    \item The \textit{Gaia} spacecraft \citep{GaiaCollab2016}. We use the \textit{Gaia}-based cluster membership probabilities and the $G$ band ($\lambda_{\mathrm{eff}} = 673$ nm) magnitudes provided by \citet{Vasiliev2021} (hereafter VB21). 
    
    \item The Two Micron All Sky Survey \citep[2MASS;][]{Skrutskie2006}. We include $J$ band ($\lambda_{\mathrm{eff}} = 1.235~\mu$m) photometry for sources not significantly affected by blending or crowding. The $J$ band is sufficient to constrain the IR tail of the SEDs for PAGB stars. $H$ and $K_s$ bands are more sensitive to variable observing conditions and do not significantly improve the SED modeling.
\end{enumerate}

Table~\ref{tab:filters} provides a summary of the filters, effective wavelengths, and bandpass widths used in this work. For the \textit{uBVI} and UVOT data, we only include objects with \texttt{DAOPHOT} goodness-of-fit parameters of $\texttt{chi} < 3$ and $-1 < \texttt{sharp} < 1$. Two bright sources with UVOT $\texttt{chi} \approx 3.3$ and $4.4$ were retained, as bright objects can exhibit elevated \texttt{chi} values when the observed profile deviates from the adopted analytical PSF model. These cuts remove most objects with poor PSF fits and sources in the clusters' densest regions, where the photometry is suspect.

\begin{deluxetable}{lccc}
\tablewidth{0pt}
\tablecaption{Filter Properties\label{tab:filters}}
\tablehead{
\colhead{Filter} & \colhead{$\lambda_{\mathrm{eff}}$} & \colhead{FWHM} & \colhead{Reference}
}
\startdata
GALEX \textit{FUV} & 1516 \AA & 234 \AA & 1, 2 \\
GALEX \textit{NUV} & 2267 \AA & 795 \AA &  \\
\hline
UVOT \textit{uvw2} & 2030 \AA & 657 \AA & 3, 4 \\ 
UVOT \textit{uvm2} & 2231 \AA & 498 \AA &  \\
UVOT \textit{uvw1} & 2634 \AA & 693 \AA &  \\
UVOT \textit{u} & 3501 \AA & 693 \AA &  \\
\hline
Thuan-Gunn $u$ & 3530 \AA & 400 \AA & 5 \\
\hline
JKC $B$ & 4405 \AA & 1008 \AA & 6 \\
JKC $V$ & 5470 \AA & 826 \AA &  \\
JKC $I$ & 8020 \AA & 1543 \AA &  \\
\hline
\textit{Gaia} $G$ & 585 nm & 440 nm & 7 \\
\hline
2MASS $J$ & 1.235 $\mu$m & 0.162 $\mu$m & 8 \\
\enddata
\tablecomments{The Thuan-Gunn $u$, Johnson-Krons-Cousins (JKC) \textit{BVI}, UVOT, \textit{Gaia} $G$, and 2MASS $J$ magnitudes are defined in the Vega system with zero-points set at a Vega magnitude of zero, except for $u$ which is set at $u=1.00$. \textit{GALEX} \textit{FUV} and \textit{NUV} are defined in the AB system.\\
References: (1) \citet{Morrissey2007}, (2) \citet{Roming2005}, (3) \citet{Poole2008}, (4) \citet{Siegel2014}, (5) \citet{ThuanGunn1976}, (6) \citet{Fukugita1996}, (7) \citet{Jordi2010}, (8) \citet{Cohen2003}.}
\end{deluxetable}

\subsection{Cluster Membership}\label{sec:membership}
\citetalias{Vasiliev2021} used \textit{Gaia} Early Data Release 3 \citep[EDR3;][]{GaiaCollab2021} parallaxes and proper motions (PMs) to derive reliable membership probabilities for stars in 170 Galactic GCs via a mixture modeling procedure. We initially used these data to select stars with membership probabilities $P > 70\%$ as our primary cluster sample, and then examined their astrometric properties to confirm membership.

\begin{figure*}[htbp]
    \centering
    \includegraphics[width=\linewidth]{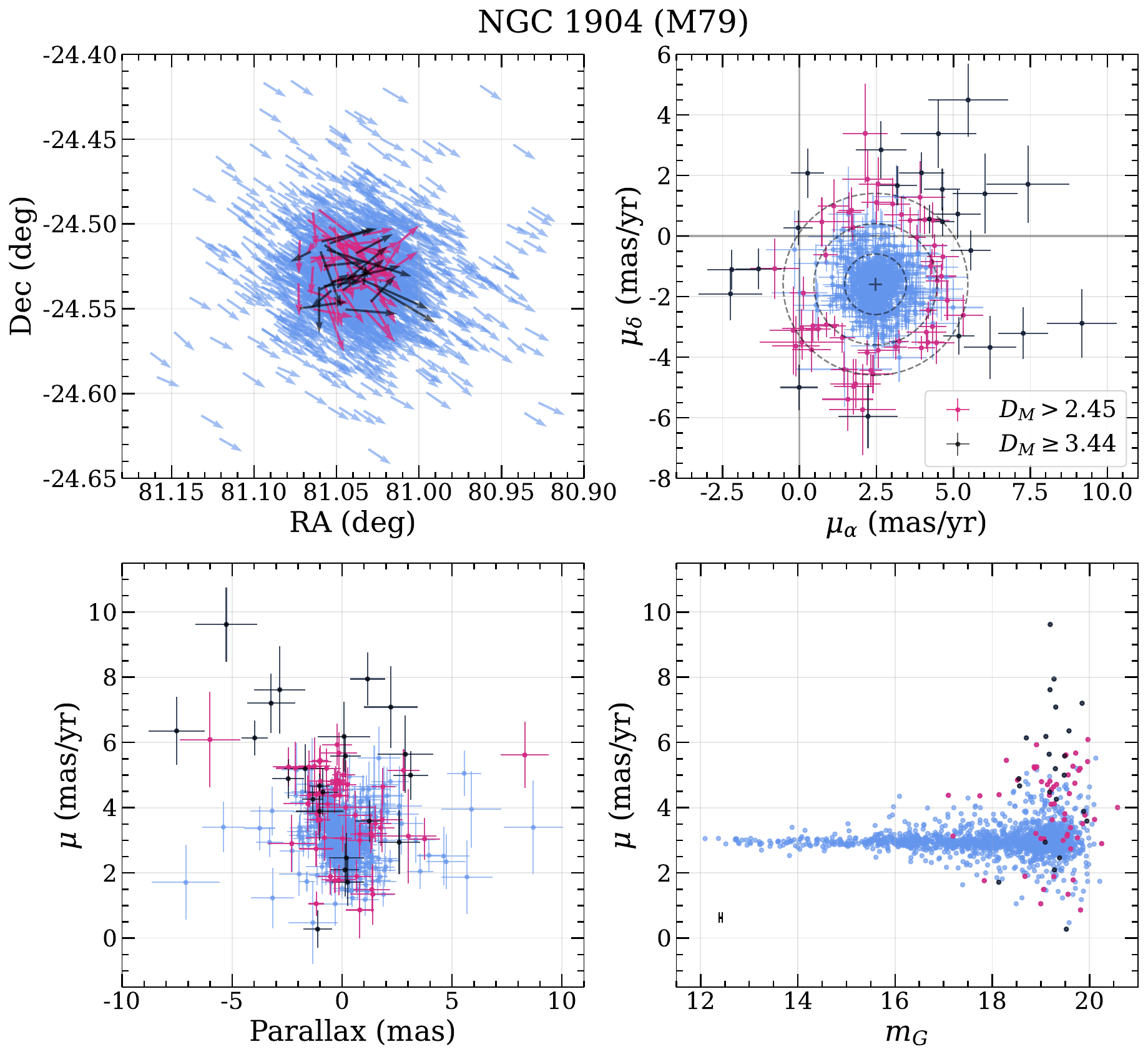}
    \caption{Astrometric diagnostics of NGC~1904 for selected cluster members with membership probabilities $P>70\%$. Top left: PM vector point diagram; objects with significant kinematic deviations from the cluster mean (identified using the Mahalanobis distance in two-dimensional PM space) are shown in pink and black for the $2\sigma$ and $3\sigma$ thresholds, respectively. Top right: PMs in right ascension ($\mu_{\alpha}$) and declination ($\mu_{\delta}$) with 1, 2, and 3 mas yr$^{-1}$ circles marking increasing distance from the cluster mean (indicated by the plus sign). Bottom left: Total PM versus EDR3 parallax; approximately 33\% of the sources have negative parallaxes, consistent with low-S/N measurements at faint magnitudes or in crowded cluster centers. Bottom right: Total PM versus \textit{Gaia} $G$-band apparent magnitude, demonstrating increasing PM uncertainties toward fainter magnitudes (a typical $G$-band magnitude error is indicated in the bottom-left corner in black). The outlying sources remain likely cluster members, as nearly all are faint ($G \sim 19\,\text{--}\,21$) and lie near the crowded cluster center where \textit{Gaia} astrometry is less reliable.}
    \label{fig:M79_membership}
\end{figure*}

Figure~\ref{fig:M79_membership} presents the astrometric diagnostics of potential cluster members for one of the GCs in our sample: NGC~1904 (M79) ---  a moderate latitude ($b = -29.35^\circ$) cluster with little reddening ($E(B-V)=0.01$). We quantified each star's kinematic deviation from the cluster mean using a Mahalanobis-like distance \citep{Mahalanobis1936},
\begin{equation}
    D_M = \sqrt{\frac{(\mu_{\alpha,*} - \langle \mu_{\alpha, \mathrm{GC}} \rangle)^2}{\sigma_{\mu_{\alpha*}}^2 + \sigma_{\mathrm{GC}}^2} + \frac{(\mu_{\delta,*} - \langle \mu_{\delta, \mathrm{GC}} \rangle)^2}{\sigma_{\mu_{\delta*}}^2 + \sigma_{\mathrm{GC}}^2}}
\end{equation}
where $\mu_{\alpha,*}$ and $\mu_{\delta,*}$ are the stellar PM components, $\langle \mu_{\alpha, \mathrm{GC}} \rangle$ and $\langle \mu_{\delta, \mathrm{GC}} \rangle$ are the adopted cluster mean PMs, $\sigma_{\mu_{\alpha*}}$ and $\sigma_{\mu_{\delta*}}$ are the PM uncertainties, and $\sigma_{\mathrm{GC}}$ is the intrinsic cluster PM dispersion, estimated from the standard deviation of the PM residuals relative to the cluster mean. Sources with $2.45 < D_M < 3.44$ were flagged as possible kinematic outliers, while those with $D_M \geq 3.44$ were identified as stronger outliers. 

These diagnostics provide an additional verification step beyond the probabilistic membership assignments from \citetalias{Vasiliev2021}. While our sample is limited to stars with membership probabilities greater than 70\%, the astrometric behavior of individual sources still requires evaluation, particularly at faint magnitudes or in crowded regions where \textit{Gaia} uncertainties increase. The PM vector field and $\mu_{\alpha}$–$\mu_{\delta}$ diagrams shown in Figure~\ref{fig:M79_membership} confirm that the majority of stars cluster tightly around the expected bulk motion of NGC~1904, with only a small number of objects showing measurable kinematic deviations from the cluster mean. These outliers, however, remain reasonable candidates for membership.  Our analysis confirms that most of these outliers are faint where astrometric precision declines ($G \sim 19\,\text{--}\,21$) and/or are located near the cluster core where crowding effects are severe. The trends in total PM versus parallax and $G$ magnitude further support this interpretation, demonstrating that scatter is strongly correlated with measurement uncertainty.  However, the PAGB and other AHB objects targeted in this work are among the brightest GC stars, and are therefore not generally affected by these limitations.

\begin{figure*}[htbp]
    \centering
    \includegraphics[width=\linewidth]{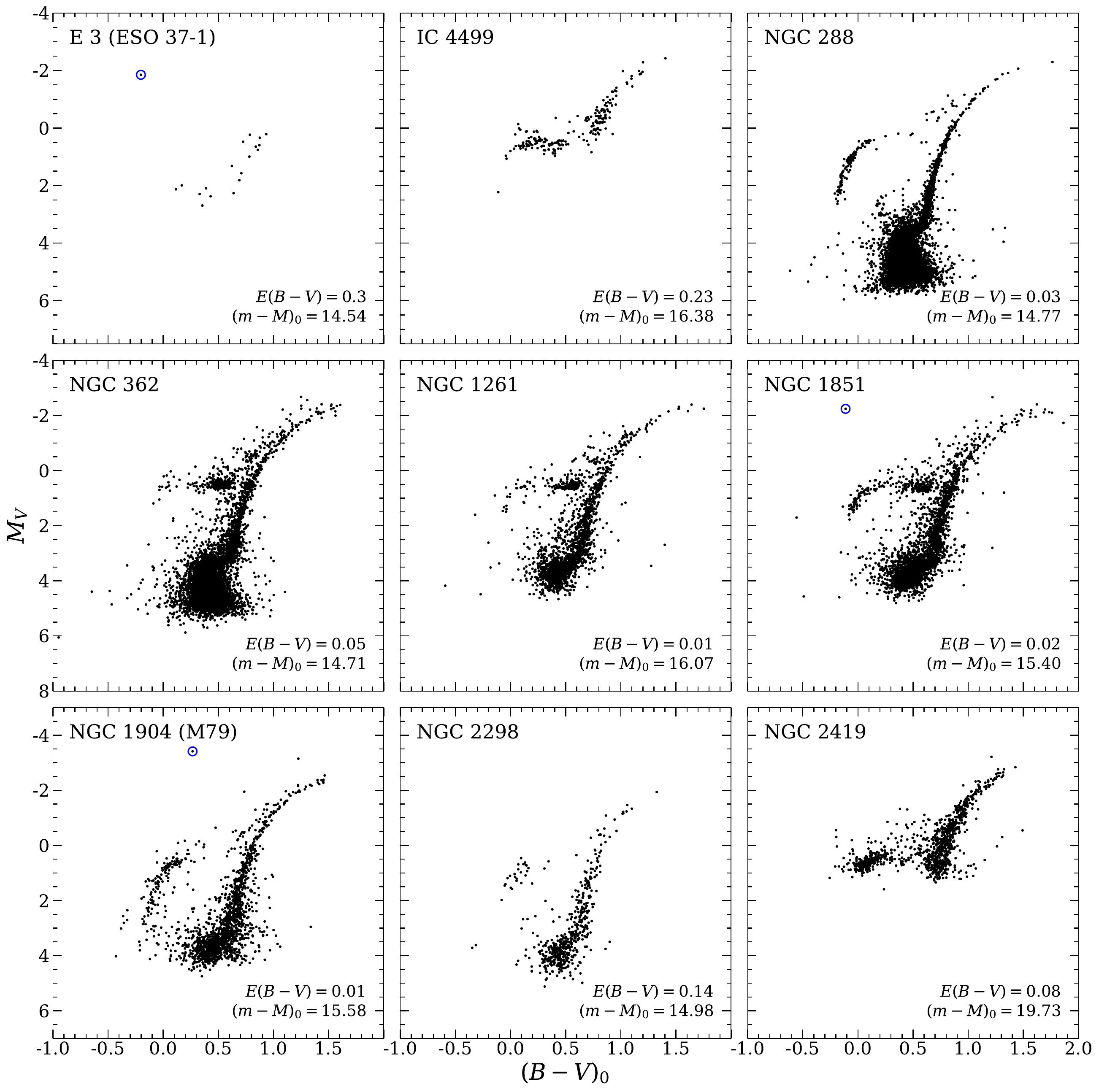}
    \caption{Optical CMDs for nine GCs evaluated in this study. Each panel shows cluster members with $-1<(B-V)_0<2$, with membership determined by the procedure discussed in Section~\ref{sec:membership}. Photometry is corrected for foreground reddening and distance using the values given in each figure and listed in Table~\ref{tab:GCParameters}. NGC~288 and NGC~1904 are representative metal-poor GCs, $\mathrm{[Fe/H]}\leq-1.28$, with well-populated BHBs; NGC~362 and NGC~1261 are metal-rich GCs, $\mathrm{[Fe/H]}>-1.28$, with defined RHBs and a red clump at $M_V \sim 0.5$. The PAGB candidates identified in Section~\ref{sec:PAGBclass} are circled in blue.}
\label{fig:CMD_panel1}
\end{figure*}

\begin{figure*}[htbp]
    \centering
    \includegraphics[width=\linewidth]{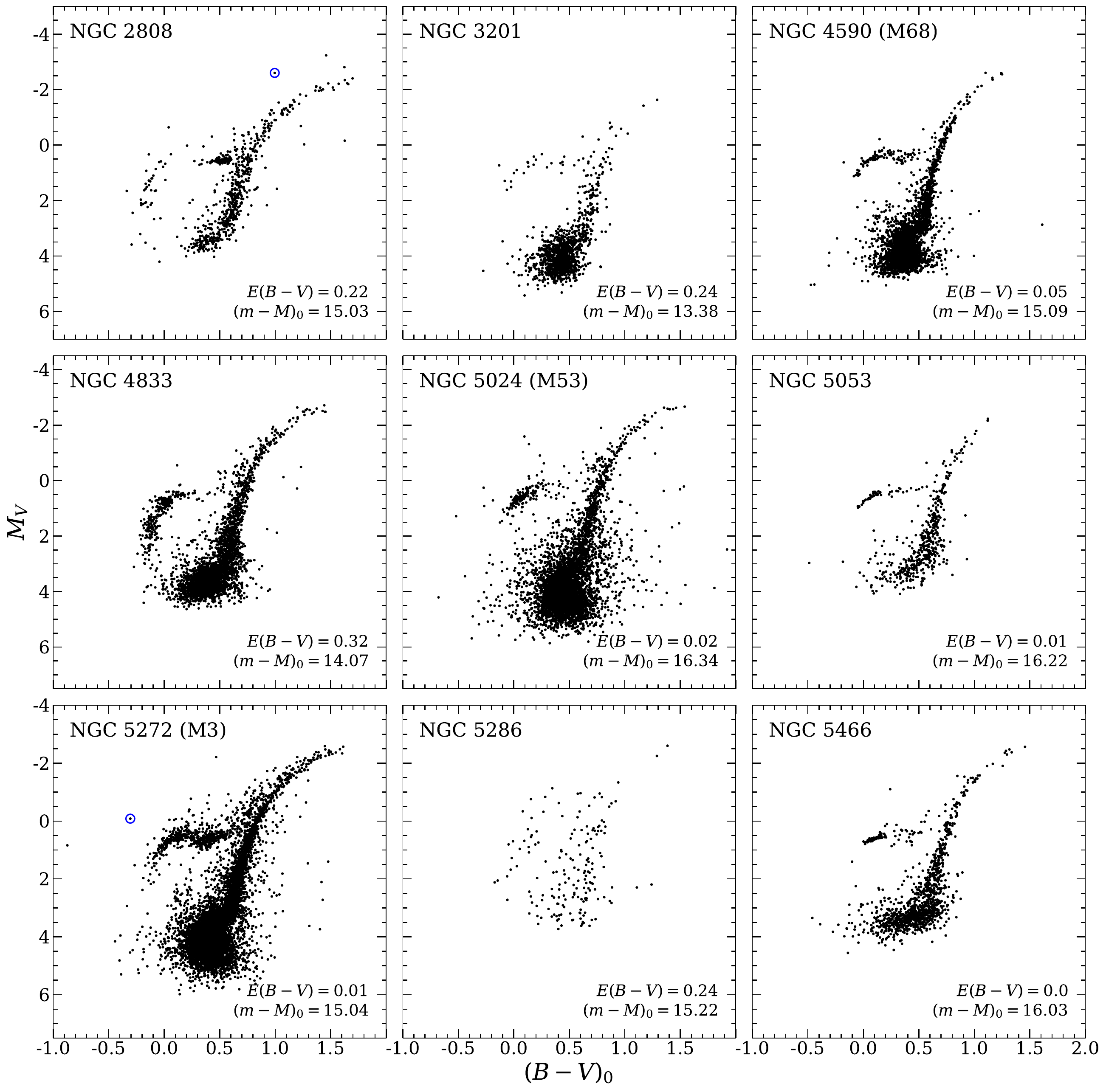}
    \caption{Same as Figure~\ref{fig:CMD_panel1} for nine GCs evaluated in this study.}
\label{fig:CMD_panel2}
\end{figure*}

\begin{figure*}[htbp]
    \centering
    \includegraphics[width=\linewidth]{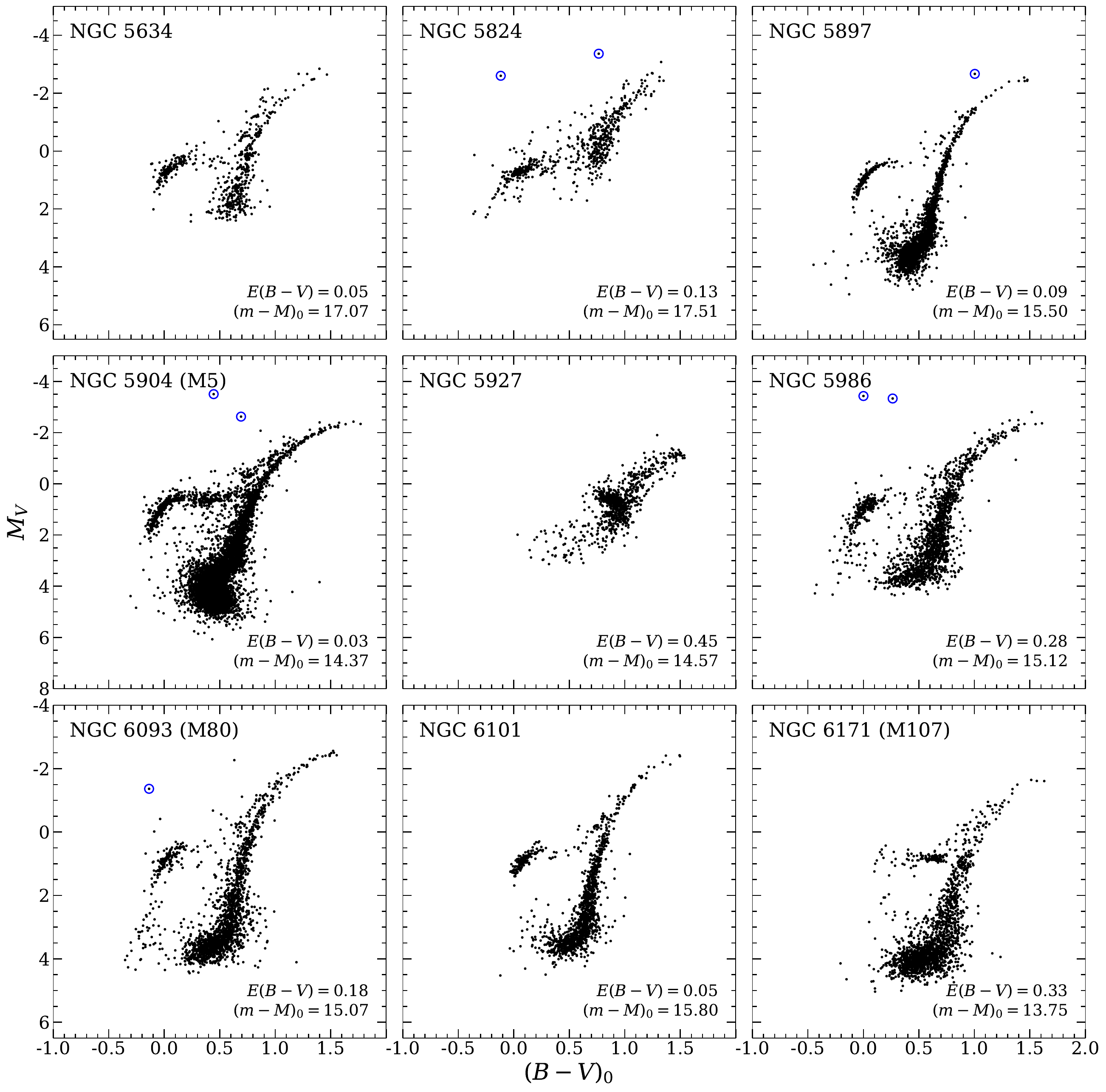}
    \caption{Same as Figure~\ref{fig:CMD_panel1} for nine GCs evaluated in this study.}
\label{fig:CMD_panel3}
\end{figure*}

\begin{figure*}[htbp]
    \centering
    \includegraphics[width=\linewidth]{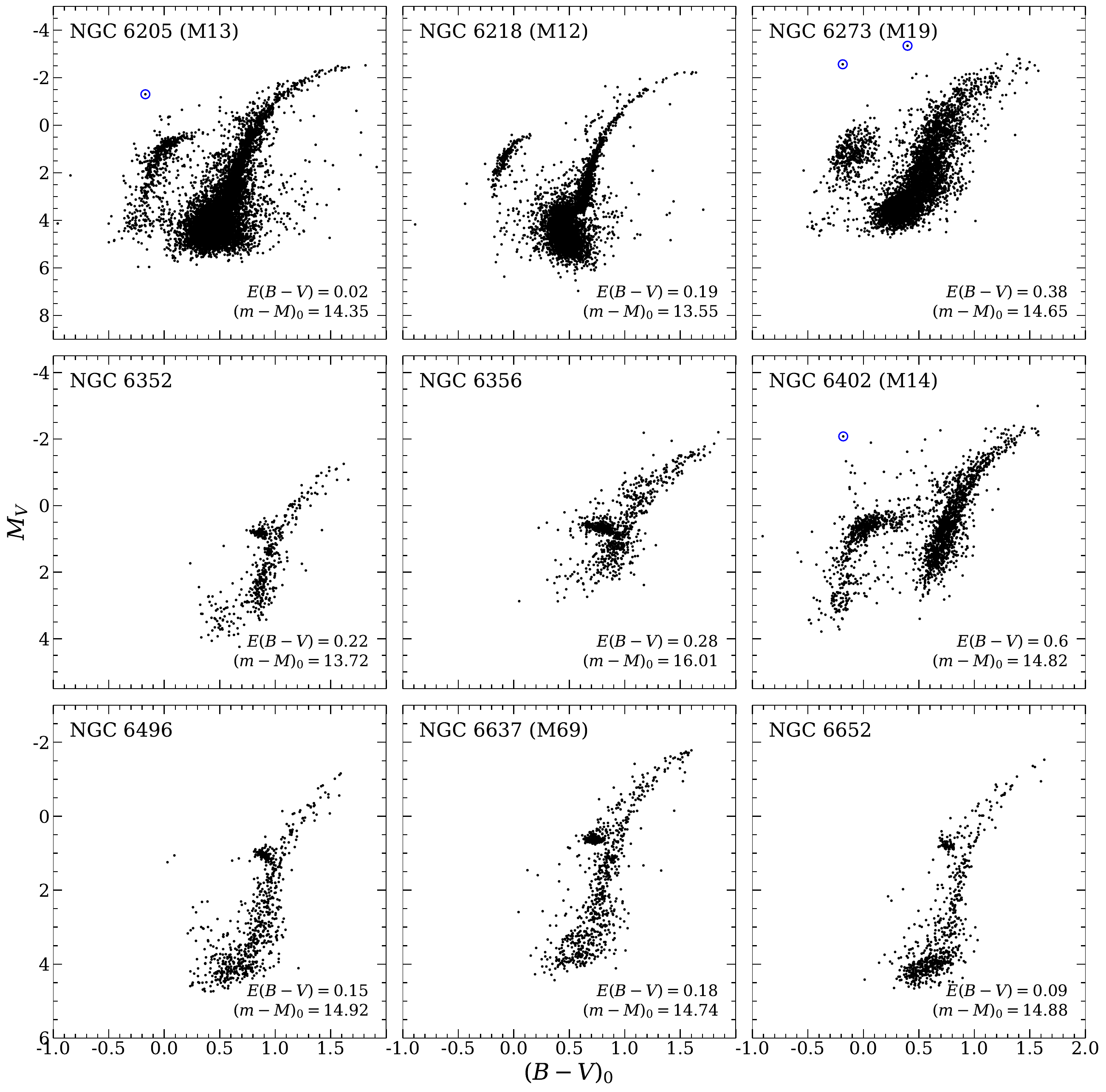}
    \caption{Same as Figure~\ref{fig:CMD_panel1} for nine GCs evaluated in this study.}
\label{fig:CMD_panel4}
\end{figure*}

Together, our diagnostic tests indicate that the final sample is a clean and reliable set of cluster members for CMD analysis and PAGB candidate identification. Although several stars appear as kinematic outliers in these diagrams, their faint magnitudes or location within the dense cluster cores suggest that their astrometry is dominated by measurement uncertainties rather than field-star contamination. Therefore, we assume these sources are cluster members. Figures~\ref{fig:CMD_panel1}--\ref{fig:CMD_panel4} present the resulting optical CMDs of identified cluster members for the GCs evaluated in this work. 

\begin{figure*}[htbp]
    \centering
    \includegraphics[width=\linewidth]{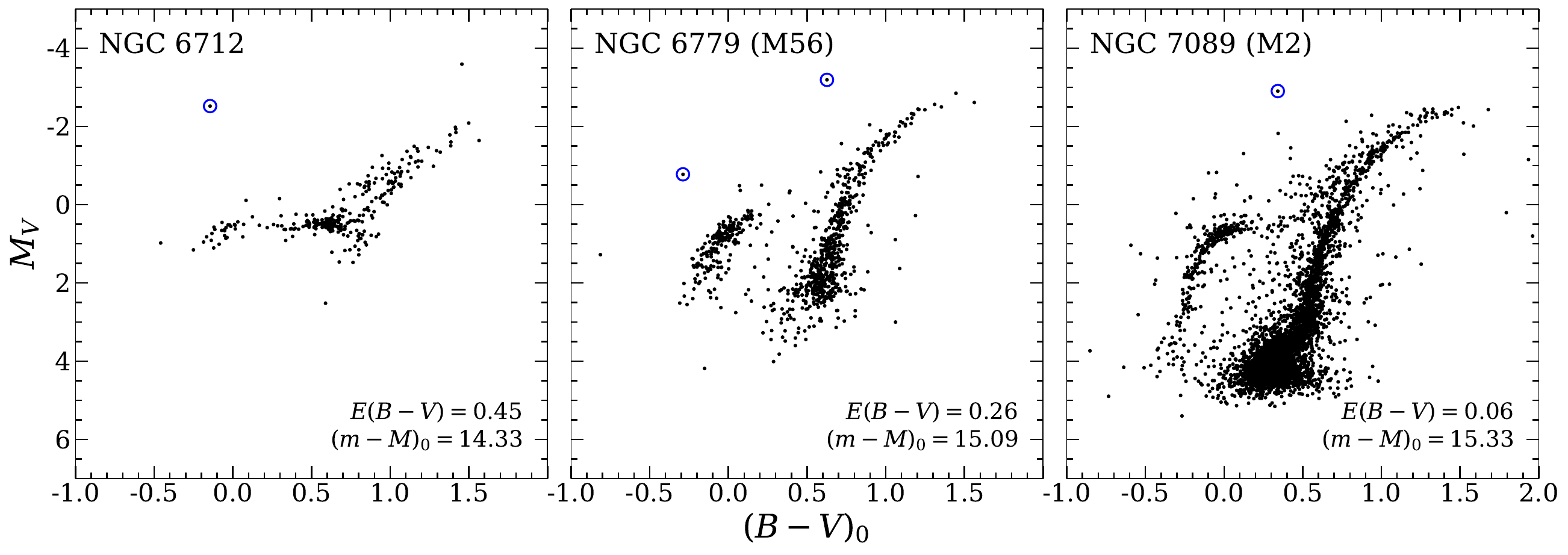}
    \caption{Same as Figure~\ref{fig:CMD_panel1} for the remaining GCs evaluated in this study.}
\label{fig:CMD_panel5}
\end{figure*}

\section{Post-AGB Stars}\label{sec:PAGBcensus}
\subsection{Post-Horizontal Branch Stellar Evolution}\label{sec:postHBevolution}
\begin{figure*}[htbp]
    \centering
    \includegraphics[width=\linewidth]{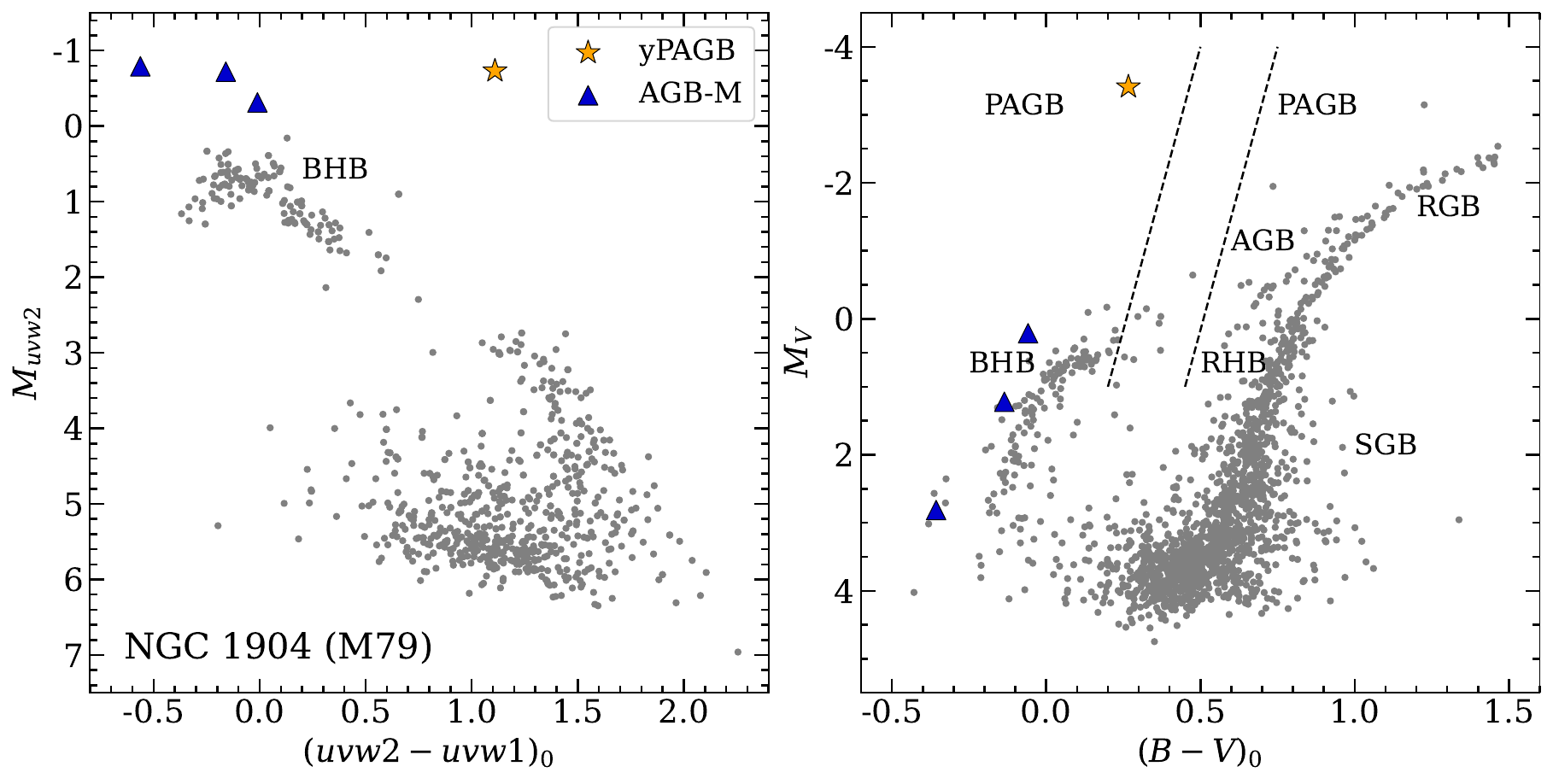}
    \caption{UV and optical CMDs for NGC~1904. Photometry is corrected for foreground reddening of $E(B-V)=0.01$ and assumes a cluster distance of 13.08 kpc. The approximate location of the instability strip is indicated by dashed lines and the stellar evolutionary stages are labeled: subgiant branch (SGB), red-giant branch (RGB), blue horizontal branch (BHB), red horizontal branch (RHB), asymptotic giant branch (AGB), and post-asymptotic giant branch (PAGB\null). AGB-Manqu\'e (AGB-M) stars and one yPAGB star are indicated by with blue triangles and yellow star marker, respectively.}
    \label{fig:CMD_M79}
\end{figure*}

At optical wavelengths, intermediate-temperature (yellow) PAGB stars are brightest objects produced by Population~II systems. This is illustrated in Figure~\ref{fig:CMD_M79}, which shows UV ($M_{uvw2}$ versus\ $uvw2-uvw1$) and optical ($M_V$ versus\ $B-V$) photometry for confirmed cluster members of NGC~1904.  The photometry has been corrected for reddening using the $E(B-V)$ value given by \citet{Harris2010} and the parameterized extinction law of \citet{Cardelli1989} (hereafter CCM89), applied over each filter bandpass assuming an average Milky Way diffuse value of $R_V = 3.1$. Approximate locations of the instability strip are indicated in the optical CMD, and positions of stars at major evolutionary stages are labeled: the subgiant branch (SGB), red-giant branch (RGB), blue horizontal branch (BHB), red horizontal branch (RHB), asymptotic giant branch (AGB), and post-asymptotic giant branch (PAGB\null).

After departing the main sequence, intermediate-mass stars ($0.8 M_{\odot} \leq M \leq 8 M_{\odot}$) will begin hydrogen shell-burning around an inert helium core. The core contracts while the envelope above the shell expands, and the star ascends the RGB\null. Once core helium burning ignites at the tip of the RGB, the star settles onto the zero-age horizontal branch (ZAHB\null). The position of the star on the ZAHB is primarily determined by its envelope mass --- the smaller the envelope mass, the bluer the star's position.

Though the relation is not perfect \citep[see for example][for a discussion of the second parameter problem]{Catelan2009, Dotter2010, Gratton2010}, the distribution of stars on the ZAHB depends primarily on the GC metallicity: metal-poor clusters tend to produce a well-populated BHB, while metal-rich clusters generate more RHB stars. As shown in Figure~\ref{fig:CMD_panel1} and Table~\ref{tab:GCParameters}, several clusters in our sample (e.g., NGC~1904, NGC~6273, and NGC~6402) have low or intermediate-metallicities ($\mathrm{[Fe/H]}\leq-1.28$), exhibit blue HBs, and are populated with bright AHB stars, including PAGB candidates. Clusters that are more metal-rich (e.g., NGC~6171, NGC~6352, and NGC~6356) have redder HBs and prominent red clumps (RCs) at $M_V \sim 0.5\,\text{--}\,1$. 

One quantitative measure of HB morphology is the horizontal branch ratio (HBR) described in \citet{Lee1994},
\begin{equation} \label{eq:HBR}
    \mathrm{HBR} = (B-R)/(B+R+V)
\end{equation}
where $B$, $V$, and $R$ are the number of blue HB stars, RR Lyrae variables, and red HB stars, respectively. An HBR greater than zero indicates a cluster with a predominantly blue HB, while a value less than zero indicates a red HB\null. Table~\ref{tab:GCParameters} includes these values (when available), along with cluster metallicity, for each system in our sample. 

Once the core helium has been exhausted, stars will undergo helium shell-burning and move toward the base of the AGB, ascending rapidly as they increases in luminosity and enter the final stages of stellar evolution. Stars in this phase are characterized by two nuclear burning shells of hydrogen and helium surrounding a carbon-oxygen core. Thermal instabilities in the helium-burning shell produce recurrent thermal pulses (TPs) during the final AGB stage. If the star has a sufficiently low envelope mass, shell burning can reach the surface before TPs begin, and the star will evolve to higher temperatures as a post-early-AGB (PEAGB) object located between the HB and PAGB regions on the CMD\null. Stars with a higher envelope mass will undergo TPs and significant mass loss through stellar winds until the envelope mass is reduced to $\sim 0.005\,M_{\odot}$. At this point, the star will begin its final journey blueward across the CMD and become a PAGB star. 

\begin{figure}[htbp]
    \centering
    \includegraphics[width=\linewidth]{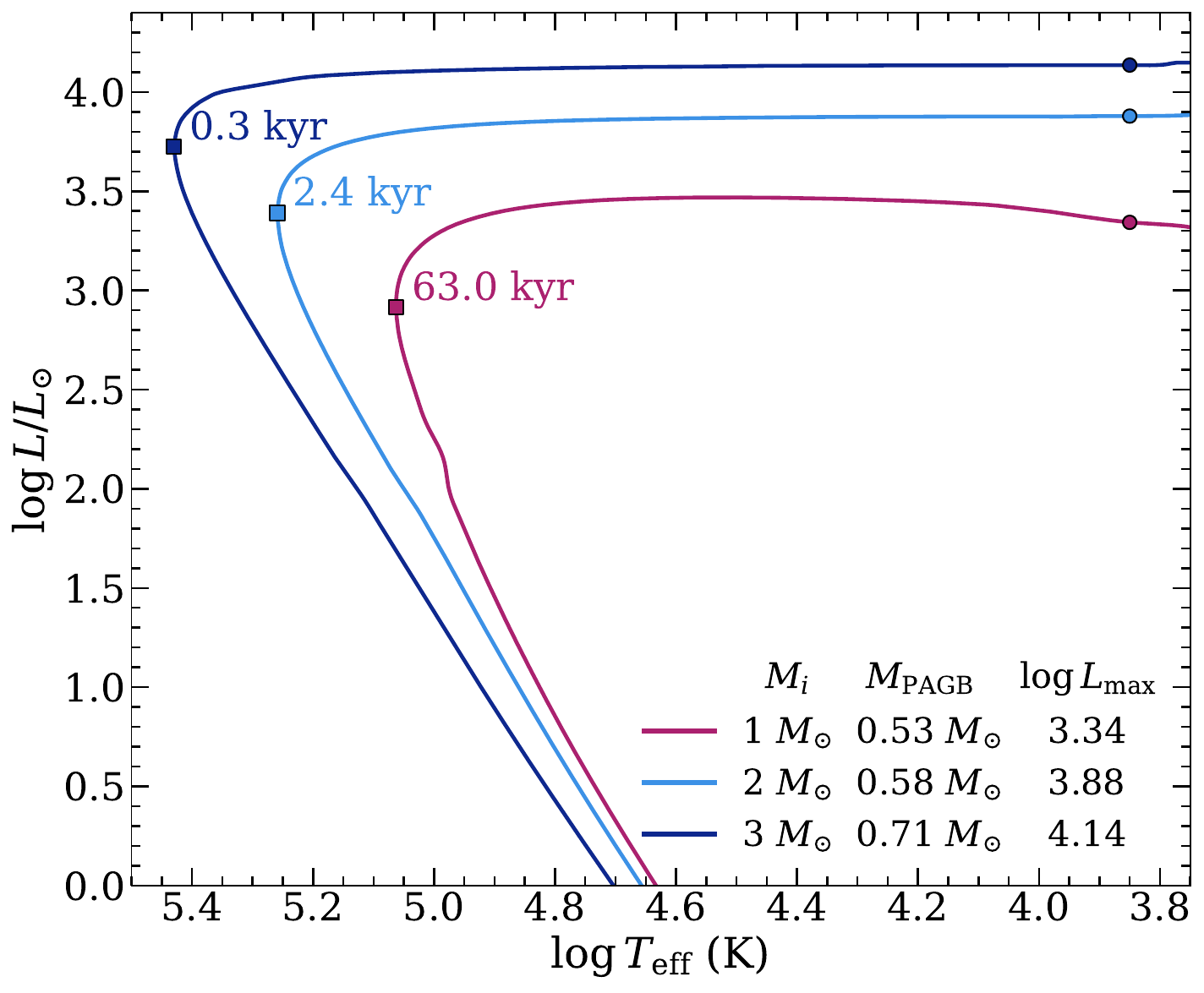}
    \caption{Post-HB evolutionary tracks from \citetalias{Bertolami2016}, illustrating the strong mass dependence of the PAGB crossing time, $\tau_{\mathrm{cross}}$. Crossing times are measured from $\log{T_{\mathrm{eff}}} = 3.85$ to the point of maximum effective temperature. The higher initial mass model ($M_i = 3~M_{\odot}$, $\log{(L_{\mathrm{max}}/L_{\odot})} = 4.14$) evolves across the HR diagram in $\tau_{\mathrm{cross}} \sim 0.3$ kyr, while the lower-mass model ($M_i = 1~M_{\odot}$, $\log{(L_{\mathrm{max}}/L_{\odot})} = 3.34$) requires $\tau_{\mathrm{cross}} \sim 62.5$ kyr.}
    \label{fig:PAGBtrack}
\end{figure}

The crossing time for PAGB stars is relatively short, strongly mass-dependent, and weakly metallicity-dependent. Models from \citetalias{Bertolami2016} show that a relatively high mass PAGB star ($\sim 0.71 M_{\odot}$) evolves rapidly, reaching its maximum effective temperature within hundreds of years and decreasing in luminosity by two orders of magnitudes in less than 1 kyr. A lower mass PAGB star ($\sim 0.53 M_{\odot}$) requires roughly 60 kyr to cross the Hertzsprung-Russell (HR) diagram and over 300 kyr to dim by two orders of magnitudes. Figure~\ref{fig:PAGBtrack} shows example evolutionary tracks from \citetalias{Bertolami2016} for varying masses (see also Figure~9 of \citetalias{Bertolami2016} for the crossing timescales across a wider range of remnant masses). Note that all PAGB objects in this study are expected to have low mass progenitors ($M_i \lesssim 1~M_{\odot}$).

Due to their rapid evolution, Population II PAGB stars are expected to be rare and exhibit a narrow luminosity function. Their specific frequency depends on both the post-HB evolutionary channel and the PAGB timescale. RHB stars --- which have comparatively massive envelopes --- generally ascend to brighter magnitudes on the AGB, and therefore produce relatively massive PAGB remnants that evolve rapidly across the HR diagram. Conversely, the most extreme BHB stars --- which have extremely low envelope masses --- may evolve as PEAGB or AGB-M objects, or otherwise never enter the classical thermal-pulsing PAGB phase. However, the slightly higher mass BHB stars that do reach the TP-AGB have less envelope mass to burn through than their RHB counterparts, and therefore end their AGB phase at comparably lower luminosities.  These lower mass cores spend much more time in the final crossing of the HR diagram. As a result, PAGB stars should preferentially be found in clusters with substantial BHB populations; most of these stars will have envelope masses large enough to enter the TP-AGB phase, but low enough to produce slow evolving PAGB objects. This is consistent with the distribution of AHB stars in the metal-poor, BHB clusters identified by \citetalias{Davis2022}.

Figure~\ref{fig:CMD_M79} shows the position of a representative GC PAGB star in the CMD.  As seen in both the UV and optical CMDs, the yPAGB star in NGC~1904 (discovered by \citet{Bond2016}) is cleanly separated from all other cluster populations, lying three magnitudes above the HB and just blueward of the instability strip. According to stellar evolution models \citep[e.g.,][]{Bertolami2016} the star will maintain a nearly constant bolometric luminosity as it evolves horizontally across the diagram.

\subsection{Candidate Selection and PAGB Classifications}\label{sec:PAGBclass}
\begin{figure*}[htbp]
    \centering
    \includegraphics[width=\linewidth]{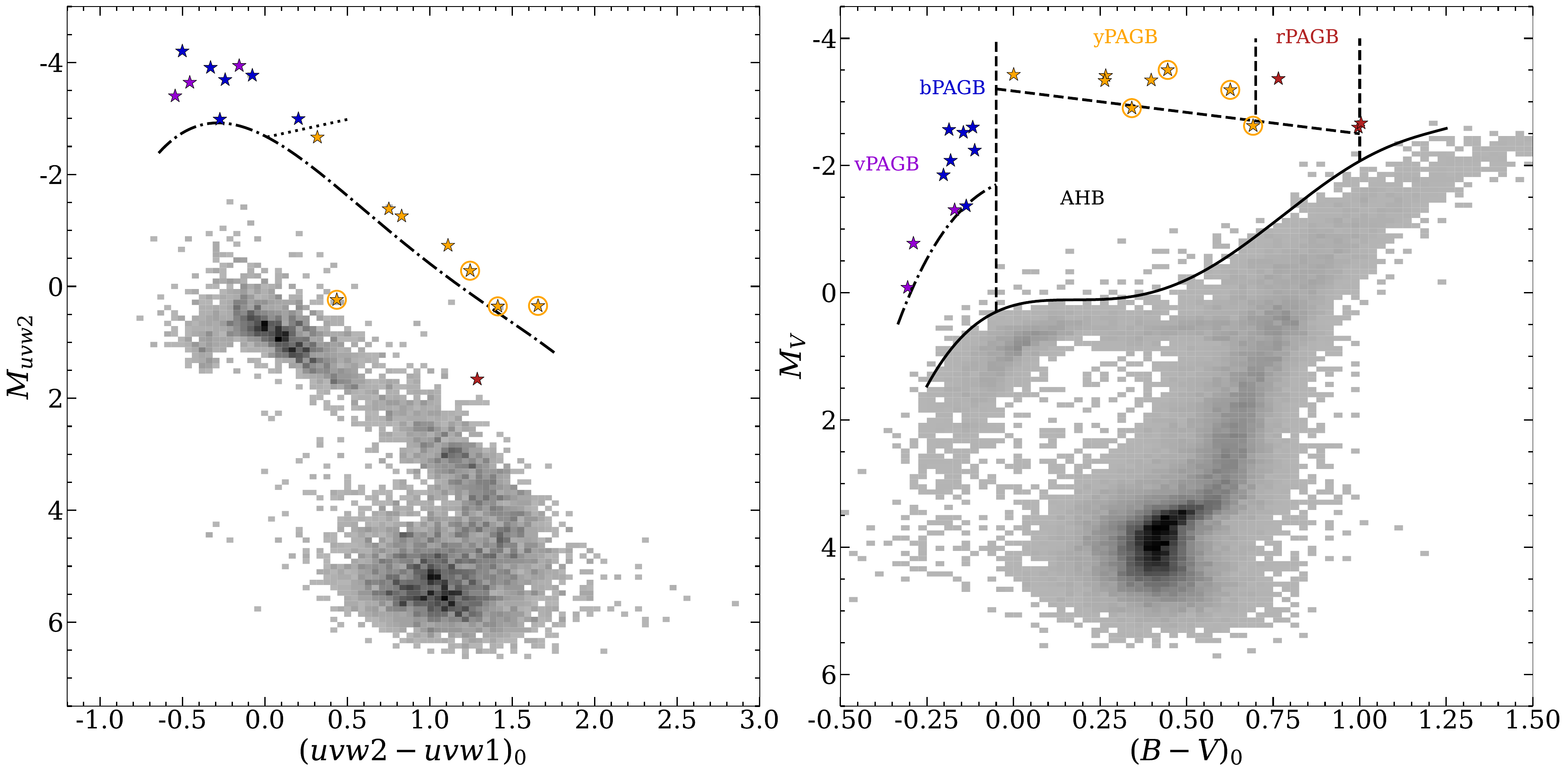}
    \caption{Stacked UV and optical Hess diagram of the GCs containing PAGB candidates. The classification regions used to identify PAGB candidates are shown. Following the definitions of \citetalias{Davis2022}, the approximate lower boundary of the AHB region is indicated by the solid line and the faint boundary of the PAGB regime is indicated by the dashed line: extending from $M_V = -3.205$ at $(B-V)_0 = -0.05$ to $M_V = -2.5$ at $(B-V)_0 = 1$. Objects with $0.7 \leq (B-V)_0 < 1.0$ are classified as red PAGB (rPAGB) stars; those with $-0.05 \leq (B-V)_0 < 0.7$ are considered yPAGBs. Blue PAGB (bPAGB) objects are hotter and occupy the region approximately two magnitudes above the HB (dash-dotted line) from $-0.25 \lesssim (B-V)_0 < -0.05$. Violet PAGB (vPAGB) stars similarly occupy the region above the HB at $(B-V)_0 \lesssim -0.25$. In the UV diagram, the dash-dotted curve is three magnitudes brighter than the upper edge of the HB stellar distribution. The approximate division between yPAGB and bPAGB/vPAGB regions are shown by the dotted line. Known RV Tauri variable stars \citep{Clement2001} are enclosed with circle markers.} \label{fig:PAGBmap}
\end{figure*}

Following the AHB classification scheme of \citetalias{Davis2022}, which builds on the \textit{Gaia}-based framework of \citet{Bond2021b}, we classify PAGB stars based on their position in the optical $M_V$ versus\ $(B-V)_0$ CMD.  We begin by defining a low-luminosity boundary for intermediate-temperature PAGBs with the line extending from $M_V = -3.205$ at $(B-V)_0 = -0.05$ to $M_V = -2.5$ at $(B-V)_0 = 1$. This line approximately traces a locus of constant bolometric luminosity. Within this region, we classify sources redder than $(B-V)_0 = 0.7$ as red PAGB (rPAGB) stars, while objects with $-0.05 \leq (B-V)_0 < 0.7$ are categorized as yellow PAGB (yPAGB) stars. Blue PAGB (bPAGB) objects are hotter and occupy the region approximately two magnitudes above the HB over the color range $-0.25 \lesssim (B-V)_0 < -0.05$. Unlike rPAGB and yPAGB stars, their Balmer-jump depends primarily on temperature rather than surface gravity. At even hotter temperatures, violet PAGB (vPAGB) stars are found above the HB at $(B-V)_0 \lesssim -0.25$. 

The location of bPAGB and vPAGB stars on optical CMDs are not cleanly defined, as the $(B-V)$ color index becomes insensitive to temperature.  As a result, stars evolving off the blue HB toward cooler temperatures (post-HB stars), as well as PAGB stars evolving from the AGB early, can occupy similar regions of the diagram. As a result, photometric classification based on optical colors alone is not sufficient. Ultraviolet photometry is needed to produce constraints on stellar temperatures and luminosities and distinguish between the evolutionary pathways.

To separate the AHB region below the PAGB boundary from the locus of HB and AGB stars, we adopt the curve from \citetalias{Davis2022} which traces the upper envelope of the HB:
\begin{eqnarray}
    M_V = &-3.74x^6 + 7.03x^5 + 6.83x^4 - 19.86x^3 \nonumber\\
    &+ 8.98x^2 - 1.51x + 0.20
\end{eqnarray}
where $x = (B-V)_0$. We applied a magnitude and color shift (typically less than $\pm 0.2$ mag) to adjust the curve above the HB for each cluster, assuming the values from \citetalias{Davis2022}. Figure~\ref{fig:PAGBmap} illustrates the approximate PAGB classification boundaries. The dashed lines in the optical diagram indicate boundaries defined by \citetalias{Davis2022} for cool and intermediate temperature PAGB stars; the dash-dotted lines outline the hotter PAGB regime investigated in this work. The dash-dotted curve in the UV diagram is three magnitudes brighter than an empirical fit to the bright edge of the stacked cluster distributions. The division between the yPAGB and bPAGB/vPAGB regions is defined by the dotted line, passing through the midpoint between the bPAGB and yPAGB candidates and perpendicular to the dash-dotted curve in CMD coordinates. (Note that the unequal scaling of the axes causes the line to appear non-perpendicular in the figure.)

Using these criteria, we identify 21 PAGB candidates across 16 GCs: E~3, NGC~1851, NGC~1904, NGC~2808, NGC~5272, NGC~5824, NGC~5904, NGC~5986, NGC~5987, NGC~6093, NGC~6205, NGC~6273, NGC~6402, NGC~6712, NGC~6779, NGC~7089. These include three rPAGB, eight yPAGB, seven bPAGB, and three vPAGB stars. Table~\ref{tab:PAGBcatalog} presents the complete catalog of PAGB candidates identified in this work. For the remaining analysis, we excluded the rPAGB stars, the four known yPAGB variable stars cataloged by \citet{Clement2001}, and one vPAGB that lacked UV photometry. The resulting sample consists of 13 non-variable, intermediate-temperature and hot PAGB stars across 11 GCs.

\begin{deluxetable*}{lclccccl}
\tabletypesize{\scriptsize}
\tablewidth{0pt}
\tablecaption{Population II Post-AGB Candidates\label{tab:PAGBcatalog}}
\tablehead{
\colhead{Cluster ID} & \colhead{Star ID$^a$} &
\colhead{Other Designations} & \colhead{Class.$^b$} &
\colhead{RA} & \colhead{Dec} & \colhead{$M_V$} &
\colhead{Notes} \\
 & & & & (J2000) & (J2000) & (mag) & 
}
\startdata
E 3 (ESO 37-1) & 358 & 2MASS 09204802-7715153 & bPAGB & 140.19999 & $-$77.25426 & $-$1.85 &  \\
NGC 1851 & 4323 & UV5 & bPAGB & 78.53583 & $-$40.05042 & $-$2.24 &  \\
NGC 1904 (M79) & 3487 & 2MASS 05241036-2429206, TYC 6479-422-1 & yPAGB & 81.04317 & $-$24.48908 & $-$3.41 &  \\
NGC 2808 & 17714 &  & rPAGB & 138.01450 & $-$64.86233 & $-$2.60 &  \\
NGC 5272 (M3) & 30107 & 2MASS 13421675+2826005, vZ 1128, ZNG 1 & vPAGB & 205.56983 & \phantom{$-$}28.43353 & $-$0.08 &  \\
NGC 5824 & 1563 & 2MASS 15035859-3304006 & rPAGB & 225.99433 & $-$33.06683 & $-$3.36 &  \\
NGC 5824 & 1464 & 2MASS 15035903-3303357 & bPAGB & 225.99601 & $-$33.05991 & $-$2.60 &  \\
NGC 5897 & 2335 & 2MASS 15173044-2100104 & rPAGB & 229.37685 & $-$21.00288 & $-$2.67 &  \\
NGC 5904 (M5) & 6774 & 2MASS 15183614+0204162, V84 & yPAGB & 229.65058 & \phantom{$-$}2.07139 & $-$3.50 & RV Tauri \\
NGC 5904 (M5) & 10926 & 2MASS 15182478+0202531, V42 & yPAGB & 229.60321 & \phantom{$-$}2.04828 & $-$2.63 & W Virginis \\
NGC 5986 & 6225 & 2MASS 15460493-3747024, ID 7 & yPAGB & 236.52059 & $-$37.78401 & $-$3.33 &  \\
NGC 5986 & 7430 & 2MASS 15460333-3746440, ID 6 & yPAGB & 236.51398 & $-$37.77893 & $-$3.43 &  \\
NGC 6093 (M80) & 4862 & 2MASS 16170420-2259201, ZNG 3 & bPAGB & 244.26753 & $-$22.98895 & $-$1.36 &  \\
NGC 6205 (M13) & 17229 & 2MASS 16413368+3626077, Barnard 29, ZNG 1 & vPAGB & 250.39033 & \phantom{$-$}36.43558 & $-$1.30 &  \\
NGC 6273 (M19) & 18767 & 2MASS 17023921-2615290, ZNG 2 & bPAGB & 255.66320 & $-$26.25817 & $-$2.56 &  \\
NGC 6273 (M19) & 24885 & 2MASS 17023520-2615236, ZNG 4 & yPAGB & 255.64666 & $-$26.25672 & $-$3.34 &  \\
NGC 6402 (M14) & 7228 & 2MASS 17373332-0314513, ZNG 1, NGC 6402-31 & bPAGB & 264.38825 & $-$3.24770 & $-$2.08 &  \\
NGC 6712 & 49 & 2MASS 18530579-0842378, ZNG 1, C49 & bPAGB & 283.27414 & $-$8.71054 & $-$2.52 &  \\
NGC 6779 (M56) & 52 & 2MASS 19163578+3011388, V6 & yPAGB & 289.14906 & \phantom{$-$}30.19409 & $-$3.19 & RV Tauri \\
NGC 6779 (M56) & 338 & 2MASS 19164129+3012484, ZNG 2 & vPAGB & 289.17201 & \phantom{$-$}30.21347 & $-$0.78 &  \\
NGC 7089 (M2) & 6074 & 2MASS 21333241-0049057, V11 & yPAGB & 323.38510 & $-$0.81819 & $-$2.90 & RV Tauri \\
\enddata
\tablecomments{
$^a$ Star ID is the listed value from the \textit{uBVI} catalog. \\
$^b$ PAGB classification described in Section~\ref{sec:PAGBclass}: red post-AGB
(rPAGB), yellow post-AGB (yPAGB), blue post-AGB (bPAGB), and violet
post-AGB (vPAGB). \\
}
\end{deluxetable*}

\section{Spectral Energy Distributions}\label{sec:SED}
To characterize the PAGB stars in our sample, we determine their stellar parameters using two fitting approaches. We first implement a traditional $\chi^2$ grid search to identify the best-fitting model and obtain initial parameter estimates. These results are then used to initialize a Bayesian analysis, which interpolates the model grids to refine parameter estimates and infer uncertainties from the posterior probability distributions.

To derive effective temperatures, surface gravities, and metal abundances, we compare our multi-wavelength photometry to a grid of ATLAS9 stellar atmosphere  models \citep{CastelliKurucz2003}. The ATLAS9 library spans an effective temperature range $3500 \leq T_{\mathrm{eff}} \leq 50{,}000~\mathrm{K}$, with grid steps of $\Delta T_{\mathrm{eff}} = 250~\mathrm{K}$ for $3500 \leq T_{\mathrm{eff}} \leq 13{,}000~\mathrm{K}$ and $\Delta T_{\mathrm{eff}} = 1000~\mathrm{K}$ for higher temperatures. The grid also includes surface gravities $0.0 \leq \log g \leq 5.0$ in steps of $\Delta \log g = 0.5$ and metal abundance ranges from $[\mathrm{M/H}] = -2.5$ to $+0.5$, with discrete grid values $[\mathrm{M/H}] = \{ -2.5,\,-2.0,\,-1.5,\,-1.0,\,-0.5,\,0.0,\,+0.2,\,+0.5 \}$. The model spectra cover the UV (1000 \AA) to IR (10 $\mu$m) wavelengths with non-uniform spacing. 

We fit these model ATLAS9 spectra to the observed far-UV through IR photometry, using between 7 and 12 bands, depending on which data were available (see Table~\ref{tab:filters}).  Model spectra were reddened following the parameterized extinction law of \citetalias{Cardelli1989}, with extinction applied across each filter bandpass assuming an average Milky Way diffuse $R_V = 3.1$ value. The majority of clusters have low to moderate reddening, $E(B-V) \lesssim 0.3$. We tested alternative extinction laws, including the prescription of \citet{Fitzpatrick1999}, and found the choice does not produce significant changes in the derived parameters. However, the effect of adopting a fixed $R_V = 3.1$ can be important. This is discussed in Appendix~\ref{app:extinction} and will be explored in future work. 

\subsection{\texorpdfstring{$\chi^2$ Grid Search}{Chi-squared Grid Search}}
We performed an initial grid search using a standard $\chi^2$ statistic. For each model, the synthetic model flux density $f_i$ in filter $i$ was computed by integrating the reddened model spectrum over the corresponding filter transmission curve, using photon or energy weighting as appropriate:
\begin{equation}
    f_i =
    \frac{
        \int F_\lambda \, 10^{-0.4 A_\lambda} \, R_i(\lambda) \, W_i(\lambda) \, d\lambda
    }{
        \int R_i(\lambda) \, W_i(\lambda) \, d\lambda
    },
\end{equation}
where $F_\lambda$ is the intrinsic model flux, $A_\lambda$ is the extinction at wavelength $\lambda$, $R_i(\lambda)$ is the response function of filter $i$, and $W_i(\lambda)$ is a weighting term equal to $\lambda$ for photon-counting detectors and unity for energy detectors.

For each model, the optimal scale factor $S$ was computed via
\begin{equation}
    S = \frac{\sum_i (y_i f_i / \sigma_i^2)}{\sum_i (f_i^2 / \sigma_i^2)} ,
\end{equation}
with variance 
\begin{equation}
    \sigma_S^2 = \left(\sum_i \frac{f_i^2}{\sigma_i^2}\right)^{-1}.
\end{equation}
Here $y_i$ and $\sigma_i$ are the observed flux density and uncertainty in filter $i$, respectively. The observed magnitudes were converted to flux densities according to their respective photometric system (AB for \textit{GALEX} and Vega for all other bands).

We evaluated the goodness-of-fit and ranked the models using the reduced $\chi^2$
\begin{equation}
\chi^2_\nu = \frac{1}{N-M}\sum_{i=1}^{N}
    \left(
        \frac{y_i - S f_i}{\sigma_i}
    \right)^2 ,
\end{equation}
where $N$ is the total number of photometric data points, $M$ is the total number of fitted model parameters (here $M = 1$, corresponding to the scale factor $S$).

From the best-fitting model and scale factor we then derive the stellar radius $R = \sqrt{S}\,d$, where $d$ is the cluster distance given in Table~\ref{tab:GCParameters}. Finally, we obtained bolometric luminosities by integrating the intrinsic (unreddened) model spectrum
\begin{equation}
    L = 4 \pi R^{2} \int_{\lambda_1}^{\lambda_2} F_{\lambda} \, d\lambda ,
\end{equation}
over the wavelength grid. 

\begin{figure*}[!tp]
    \centering
    \includegraphics[width=\linewidth]{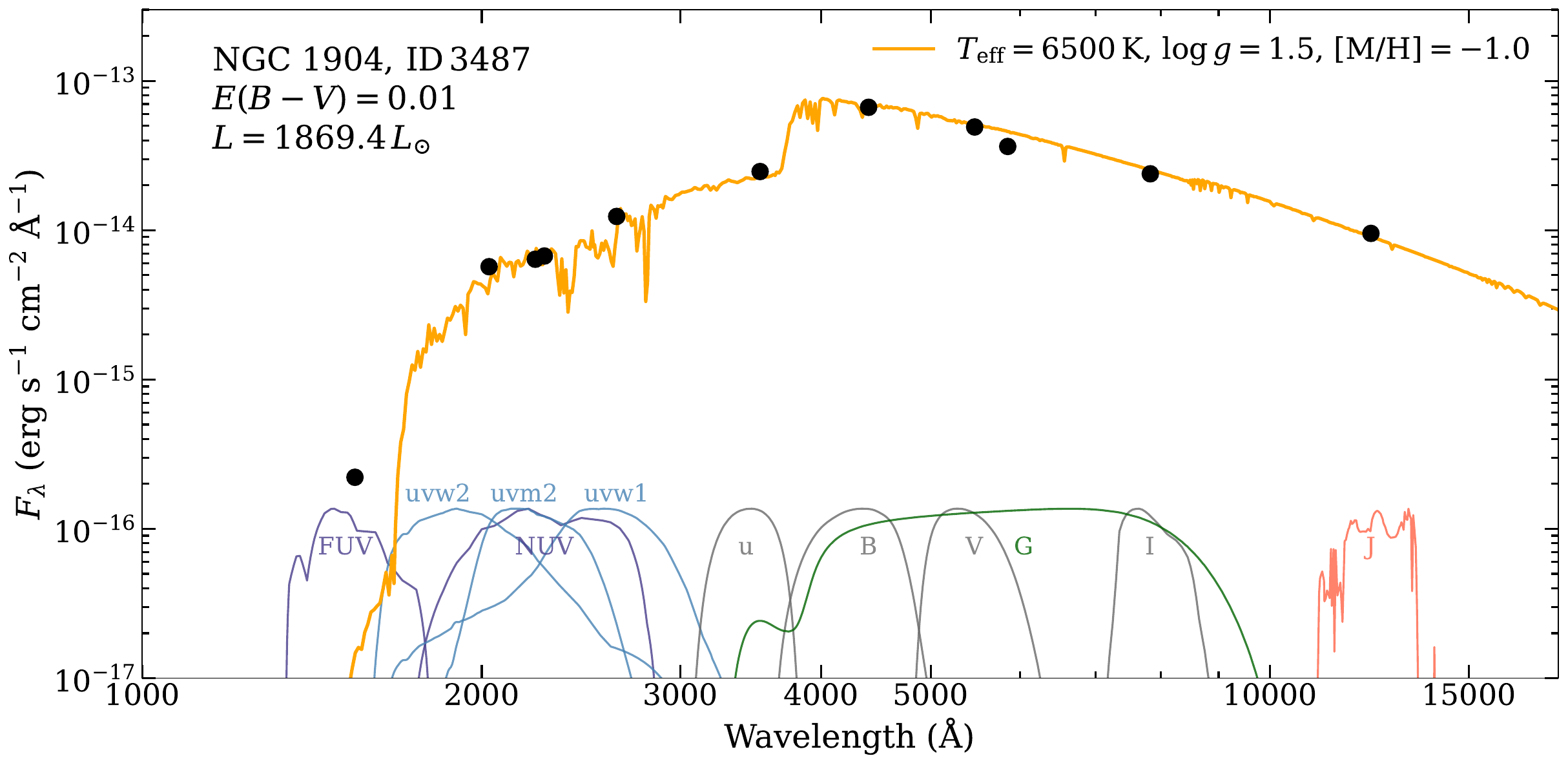}
    \caption{SED fit for the yPAGB star in NGC~1904. Black circles show the observed (reddened) photometric fluxes with uncertainties (most are smaller than the plot markers). The solid yellow curve represents the best-fit stellar atmosphere model reddened using the \citetalias{Cardelli1989} extinction law and the cluster $E(B-V)$ (labeled on the plot and listed in Table~\ref{tab:GCParameters}). The transmission curves of the filters listed in Table~\ref{tab:filters} are shown at the bottom.}
    \label{fig:SEDfit_NGC1904}
\end{figure*}

\begin{figure*}[!tp]
    \centering
    \includegraphics[width=\linewidth]{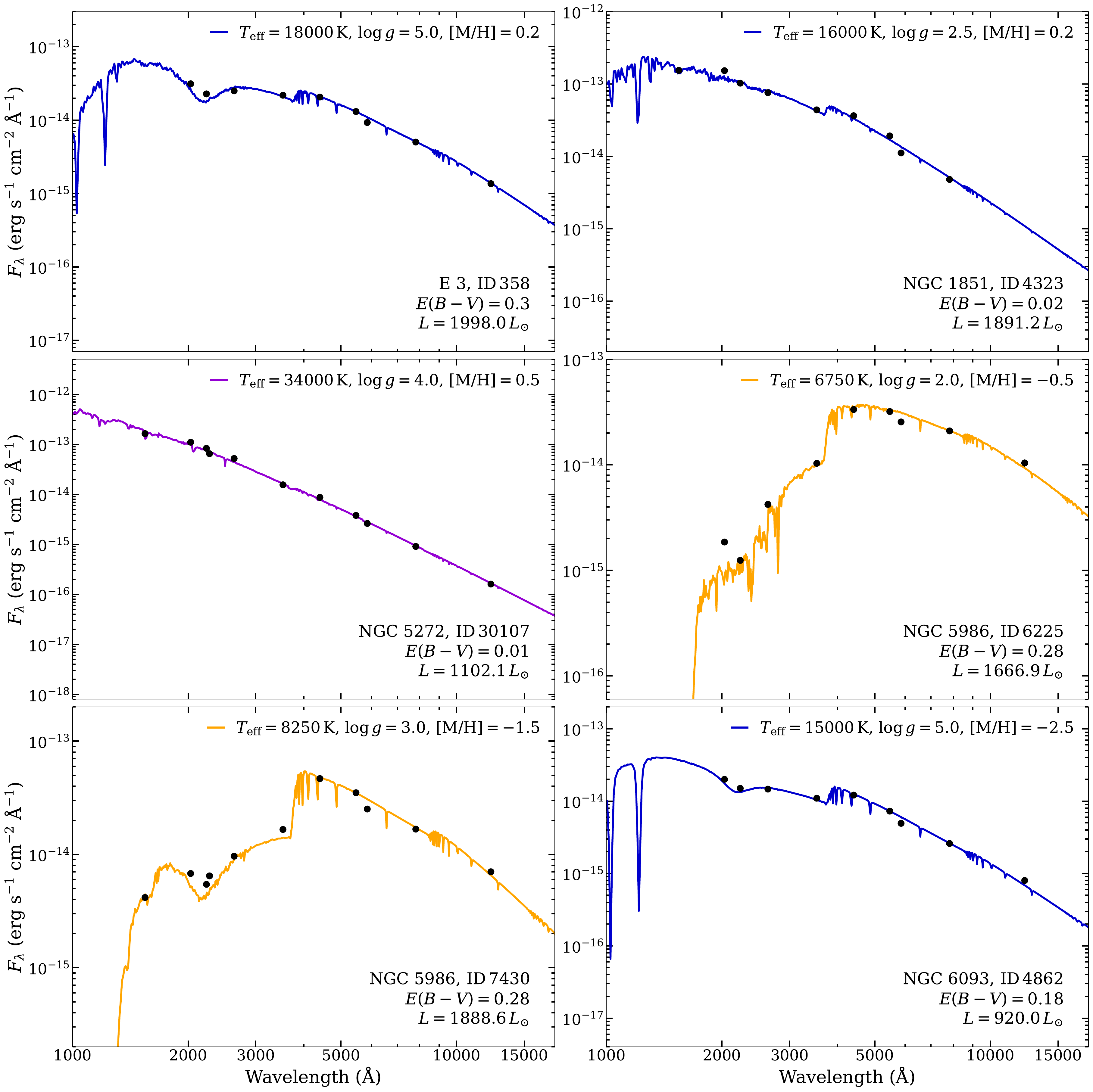}
    \caption{SED fits for the PAGB stars in this study that have complete UV--IR photometric coverage. Panels show individual SEDs, with cluster information and fitting parameters labeled on each plot. Black circles show the observed (reddened) photometric fluxes with uncertainties (most of which are smaller than the plot markers). Solid curves represent the best-fit stellar atmosphere model reddened using the \citetalias{Cardelli1989} extinction law and the cluster $E(B-V)$ (labeled on the plot and listed in Table~\ref{tab:GCParameters}). The 2175\,\AA~UV absorption feature is prominent for moderately-reddened objects, such as the PAGB stars in E 3, NGC~5986, and NGC~6093.}
\label{fig:SEDfit_PAGBs1}
\end{figure*}

\begin{figure*}[!tp]
    \centering
    \includegraphics[width=\linewidth]{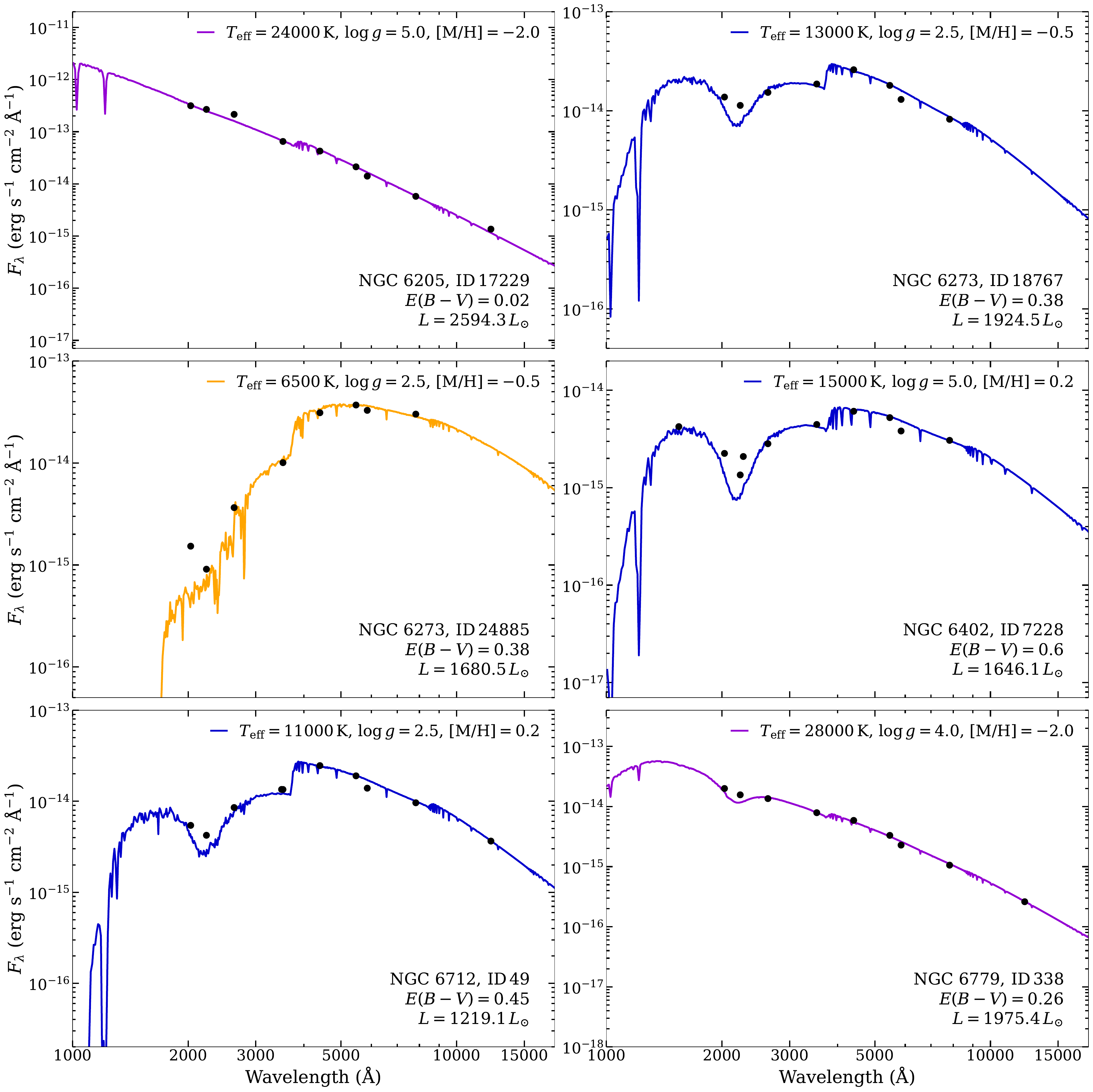}
    \caption{Same as Figure~\ref{fig:SEDfit_PAGBs1} for the remaining PAGB stars in this study.}
\label{fig:SEDfit_PAGBs2}
\end{figure*}

Figures~\ref{fig:SEDfit_NGC1904},~\ref{fig:SEDfit_PAGBs1}, and~\ref{fig:SEDfit_PAGBs2} show the resulting SED fits for the PAGB candidates identified in Section~\ref{sec:PAGBclass} that have full UV--IR photometric coverage. Fluxes are plotted at the effective wavelength of each filter.

The yPAGB stars have temperatures of $T_{\mathrm{eff}} \approx 6500\,\text{--}\,8250$ K, and luminosities of $1.5\,\text{--}\,1.9\times10^3~L_{\odot}$, while the bPAGB stars are significantly hotter, with $T_{\mathrm{eff}} \approx 12{,}000\,\text{--}\,18{,}000$ K and $L\approx 1.6\,\text{--}\,2.0\times10^3~L_{\odot}$. The vPAGB stars are even hotter, with flux densities peaking in the far-UV\null.  Thus for these objects, we can only derive lower limits for the temperature ($T_{\mathrm{eff}} > 25{,}000$ K) and luminosity ($L\approx 1.1\,\text{--}\,1.9\times10^3~L_{\odot}$).

The optical and near-IR photometry are in good agreement with model predictions. The largest residuals occur at UV wavelengths where extinction effects are the strongest and uncertainties in the adopted reddening law have a greater impact, especially for clusters with higher $E(B-V)$. This behavior is further discussed in Section~\ref{sec:limits} and Appendix~\ref{app:extinction}\null. Some discrepancies may also arise from instrumental effects, including the known red-leak in the UVOT filters, discussed in Appendix~\ref{app:UVOTfilters}\null.

At longer wavelengths, a few objects exhibit an apparent excess in IR flux relative to the models. While this could suggest the existence of emission from circumstellar dust, it is more likely due to blending of 2MASS photometry with nearby sources. This can be particularly problematic for inner regions of GCs, where crowding is severe, and the 2MASS survey resolution is a limiting factor.  \citet{Bond2021a} noted this issue for the yPAGB star in NGC~6273, which was blended with four nearby red giants. For these cases, we exclude 2MASS photometry from our fits. Overall, there is no clear evidence of circumstellar dust around any Population II yPAGB star; this is consistent with theoretical expectations that AGB dust is dissipated over the long evolutionary timescales of low-mass PAGB stars \citep{Bond2016, Bond2020}.

We examined the highest-ranked models for each object and find that the PAGB stars exhibit little scatter in the inferred temperatures and luminosities --- $\langle \sigma_{T_{\mathrm{eff}}}\rangle \approx 170$ K and $\langle \sigma_{L_{\mathrm{bol}}}\rangle \approx 65~L_{\odot}$ --- indicating these properties are well defined by the photometry and SED fits. In contrast, the PAGB stars' surface gravities and abundances are poorly constrained, with models typically spanning several grid steps ($\approx 0.4$ dex) for both.

The best-fit models visually reproduce the observed fluxes, and the consistency in the inferred temperatures and luminosities across multiple top-ranked solutions reinforces that the PAGB SEDs are successfully modeled. Overall, the SED fitting suggests that temperature and luminosity are robustly determined with broadband photometry for all PAGB stars, but the inferred abundance and surface gravities require additional evaluation.

\subsubsection{Metal Abundance and Surface Gravities}\label{sec:M-logg} 
The grid of ATLAS9 atmospheric models are defined in terms of a scaled-solar total metal abundance ([M/H]) and not metallicity ([Fe/H\null]). For old, $\alpha$-enhanced systems like GCs, [M/H] can differ from the reported cluster metallicities by
\begin{equation}
    \mathrm{[M/H]} \cong \mathrm{[Fe/H]} + \log{(0.638 f_{\alpha} + 0.632)}
\end{equation}
resulting in a $\approx 0.2\,\text{--}\,0.4$ dex offset \citep{Salaris1993}.  When considering this issue, the fitted [M/H] values are broadly consistent with the known cluster metallicities in Table \ref{tab:GCParameters}, typically within one to two model grid steps.

While GC metallicities are generally treated as chemically homogeneous, studies reveal a small but measurable intrinsic spread of $\sim 0.05\,\text{--}\,0.1$ dex that is more prominent for massive clusters \citep[e.g.,][]{Carretta2009}. The stars of NGC~6273, in particular, have been shown to possess a wide range of iron content ($\mathrm{[Fe/H]} = -2$ to $-1$) \citep{Johnson2017}. Furthermore, because PAGB surface abundances can be modified by dredge-up events, CNO processing, and chemical mixing, their present-day abundance may differ from the metallicity of the cluster's less-evolved stars or the star's natal abundance \citep{Iben1983, Herwig2005, Gonzalez1994, Dixon2024}. Because of these effects, the model fit parameter [M/H] should be considered approximate and not interpreted as a precise measure of chemical composition.

Combining the relations $L = 4\pi \sigma R^2 T_{\mathrm{eff}}$ and $g = GM/R^2$, stellar mass can be estimated by
\begin{equation}
    \log{\left(\frac{M}{M_{\odot}}\right)} = \log{\left(\frac{L}{L_{\odot}}\right)} + \log{g} - 4\log{T_{\mathrm{eff}}} + 10.61
\end{equation}
However, applying the derived physical parameters from the SED fits ($L$, $T_{\mathrm{eff}}$, and $ \log{g}$) yields a wide range of mass estimates, many of which are not physically reasonable for evolved PAGB stars: up to $15.1~M_{\odot}$ for yPAGB stars and $132.5~M_{\odot}$ for bPAGB stars. This suggests the best-fit model surface gravities are not reliably determined by the broadband photometry and resulting SED fits. Additionally, the model atmosphere grids impose discrete sampling for temperature, metallicity, and surface gravity that limits the sensitivity of derived properties. To overcome these obstacles and attempt to break degeneracies between inferred parameters, we interpolated the grid models and employed a Bayesian analysis.

\subsection{Bayesian Inference}\label{sec:Bayes} 
\begin{figure*}[!tp]
    \centering
    \gridline{
    \fig{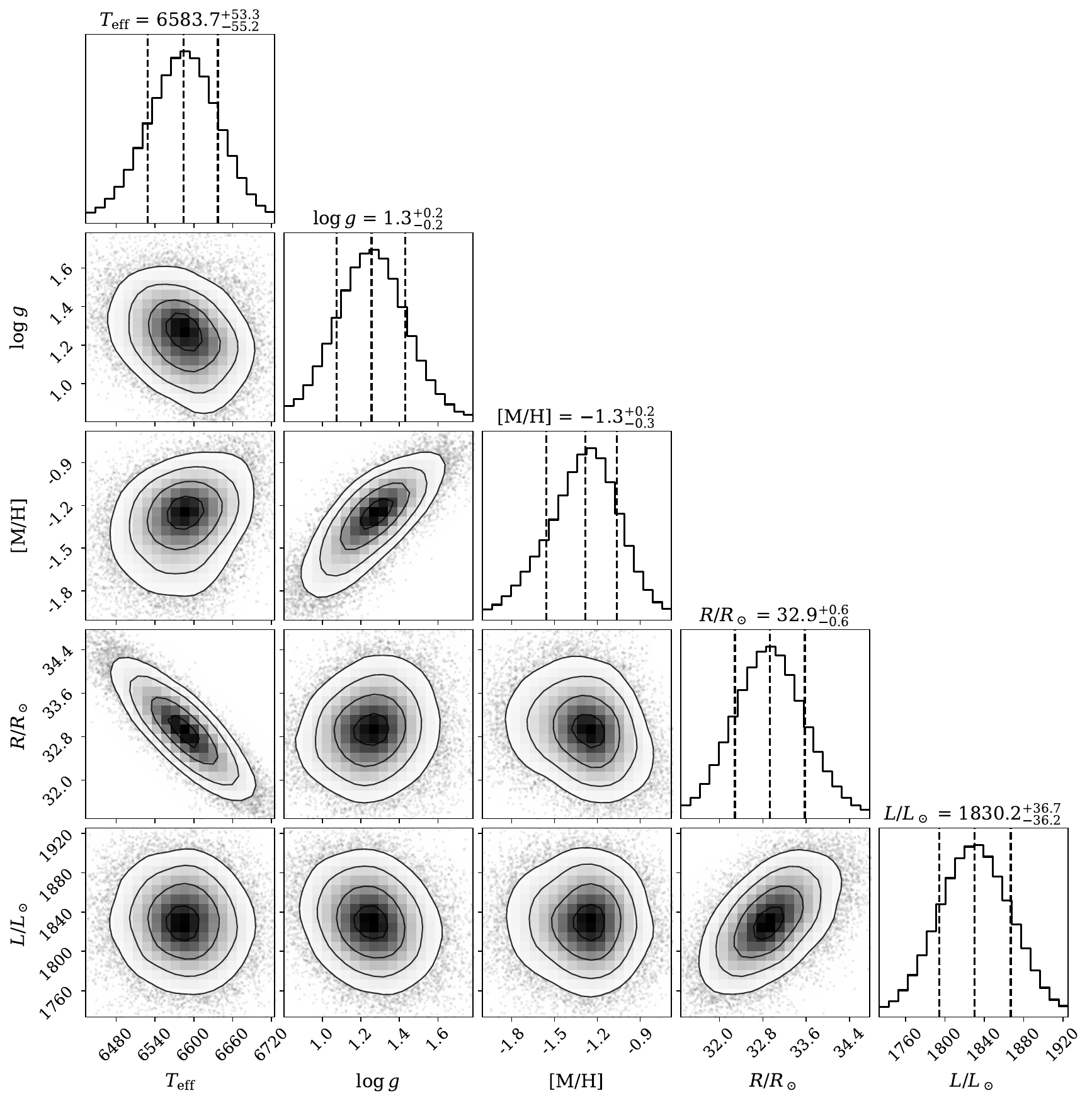}{0.49\textwidth}{(a) NGC~1904 ID 3487 (yPAGB)}
    \fig{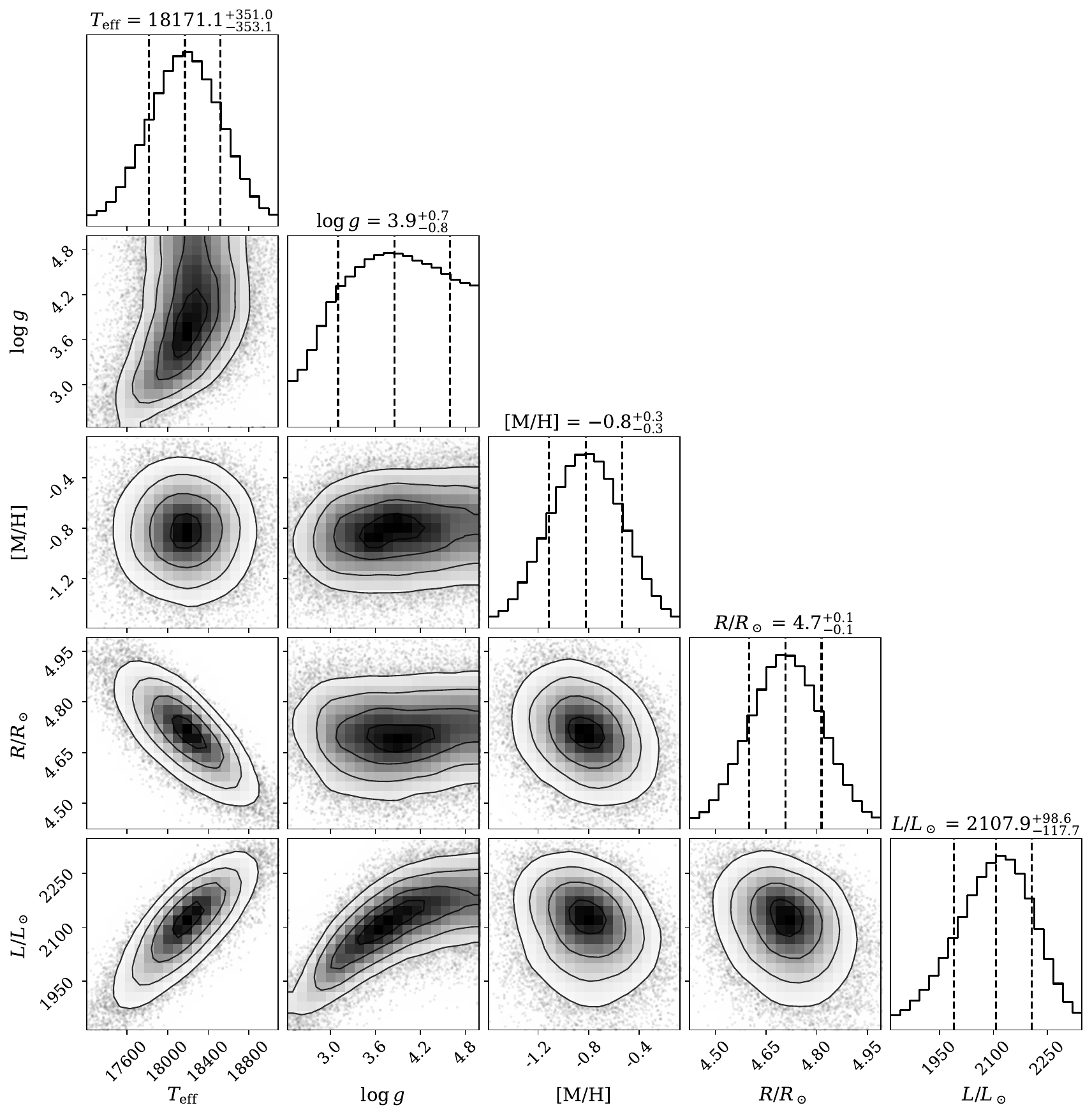}{0.49\textwidth}{(b) E3 ID 358 (bPAGB)}
        }
    \caption{Corner plots showing the posterior distributions of $T_{\mathrm{eff}}$, $\log g$, [M/H], $R$, and $L$ for the yPAGB star in NGC~1904 (left) and the bPAGB star in E3 (right). The corresponding best-fit $\chi^2$ SED fits are shown in Figures~\ref{fig:SEDfit_NGC1904} and \ref{fig:SEDfit_PAGBs1} (top-left panel), respectively. Posterior samples were generated using the MCMC analysis described in Section~\ref{sec:Bayes}. Contours and histograms show the median and 68\% credible intervals.}
    \label{fig:BayesSolution}
\end{figure*}

An alternative way of deriving the PAGB stars' physical parameters --- effective temperature $T_{\mathrm{eff}}$, surface gravity $\log{g}$, metal abundance [M/H], and stellar radius $R$ --- is through the use of a Bayesian framework. To accomplish this, we linearly interpolated the reddened ATLAS9 atmosphere grids in logarithmic flux to produce a continuous model parameter space.  Assuming Gaussian measurement uncertainties, the likelihood for a given set of model parameters $\theta = \{\log{T_{\mathrm{eff}}}, \, \log{g}, \, \mathrm{[M/H]}, \, \log{S}\}$ fitting our stellar measurements is
\begin{equation}
    \mathcal{L}(\theta) = \prod_{i=1}^N \frac{1}{\sqrt{2 \pi}\sigma_{\mathrm{o},i}} \exp{\left[
        -\chi_i^2/2\right]}
\end{equation}
where $N$ is the number of photometric bands and 
\begin{equation}
    \chi_i^2 = \left(\frac{F_{\mathrm{o},i} - SF_{\mathrm{m},i}}{\sigma_{\mathrm{o},i}}\right)^2
\end{equation}
for the scale factor $S$, observed fluxes $F_{\mathrm{o},i}$, model fluxes $F_{\mathrm{m},i}$, and observed flux uncertainties $\sigma_{\mathrm{o},i}$.

We adopt a log-uniform prior on $T_{\mathrm{eff}}$, a Gaussian prior on [M/H], centered on the cluster metallicity listed in Table~\ref{tab:GCParameters}, and uniform priors on $\log{g}$ and $\log{S}$ over the allowed model range. The log-likelihood is then
\begin{equation}
    \ln{\mathcal{L}(\theta)} = -\frac{1}{2}\sum_{i=1}^N{\left[\chi_i^2 - \ln{(2\pi \sigma_i^2))}\right]}.
\end{equation}

The posterior probability distribution was sampled using the Markov Chain Monte Carlo (MCMC) sampler implemented in \texttt{emcee} \citep{emcee}. Walkers were initialized near the best-fit $\chi^2$ solution to improve convergence and reduce secondary low-probability modes, and posterior samples were used to derive credible intervals for all fitted parameters. Stellar radii were again computed using the scaling parameter ($R = d\sqrt{S}$) and luminosities were derived from each posterior sample and corresponding bolometric model fluxes.

Figure~\ref{fig:BayesSolution} shows corner plots with the posterior distributions of $T_{\mathrm{eff}}$, $\log{g}$, [M/H], and $R$ for the yPAGB star in NGC~1904 and bPAGB star in E3. The best-fit $\chi^2$ solutions for these stars are shown in Figures~\ref{fig:SEDfit_NGC1904} and~\ref{fig:SEDfit_PAGBs1} (top-left panel), respectively. The Bayesian solutions are broadly consistent with the $\chi^2$ result: for the yPAGB star, $L = 1830 ^{+37}_{-36}~L_{\odot}$, $T_{\mathrm{eff}} = 6584^{+53}_{-55}~\text{K}$, $\log{g} = 1.3 \pm 0.2$, $\mathrm{[M/H]} = -1.3^{+0.2}_{-0.3}$, and $R = 32.9 \pm 0.6 R_{\odot}$, while for the bPAGB, $L = 2108^{+99}_{-118}~L_{\odot}$, $T_{\mathrm{eff}} = 18171^{+351}_{-353}~\text{K}$, $\log{g} = 3.9^{+0.7}_{-0.8}$, $\mathrm{[M/H]} = -0.8\pm0.3$, and $R = 4.7 \pm 0.1 R_{\odot}$. All parameters agree within the uncertainties.

Across the entire sample, the Bayesian and $\chi^2$ solutions typically agree within 6\%  in luminosity and $\sim 500~\text{K}$ in effective temperature. However, differences in metal abundances between the fits were frequently beyond the uncertainties. All Bayesian solutions favor lower abundances ($\mathrm{[M/H]} < 0$) consistent with the clusters' low-metallicity environments.  This result is most likely influenced by the adopted priors.

For the purposes of constructing the PAGB luminosity function, we adopt the luminosities derived from the Bayesian analysis. This provides a full posterior distribution for each star, and incorporates both measurement uncertainties and prior information. Table~\ref{tab:PAGBparams} presents the stellar parameters and derived properties for the final PAGB sample. 

\begin{deluxetable*}{lccccccccccc}
\tabletypesize{\scriptsize}
\tablewidth{0pt} 
\tablecaption{Derived Physical Parameters of Post-AGB Candidates \label{tab:PAGBparams}}
\tablehead{
\colhead{Cluster ID} & \colhead{Star ID$^a$} & \colhead{Class.$^b$} & \colhead{RA (J2000)} & \colhead{Dec (J2000)} &
\colhead{$T_{\mathrm{eff}}$} & \colhead{$L$} & \colhead{$R$} &
\colhead{$\log g$} & \colhead{[M/H]} \\
\colhead{} & \colhead{} & \colhead{} & \colhead{(deg)} & \colhead{(deg)} &
\colhead{(K)} & \colhead{($L_\odot$)} & \colhead{($R_\odot$)} & 
\colhead{(dex)} & \colhead{(dex)}
}
\startdata
E 3 (ESO 37-1) & 358 & bPAGB & 140.19999 & $-$77.25426 & $18171_{-353}^{+351}$ & $2107.9_{-117.7}^{+98.6}$ & $4.71 \pm 0.11$ & $3.9_{-0.8}^{+0.7}$ & $-0.8 \pm 0.3$ \\
NGC 1851 & 4323 & bPAGB & 78.53583 & $-$40.05042 & $18407_{-801}^{+952}$ & $2637.0_{-216.3}^{+193.8}$ & $5.45 \pm 0.22$ & $2.4 \pm 0.1$ & $-1.4 \pm 0.3$ \\
NGC 1904 (M79) & 3487 & yPAGB & 81.04317 & $-$24.48908 & $6584_{-55}^{+53}$ & $1830.2_{-36.2}^{+36.7}$ & $32.93_{-0.64}^{+0.65}$ & $1.3 \pm 0.2$ & $-1.3_{-0.3}^{+0.2}$ \\
NGC 5272 (M3) & 30107 & vPAGB & 205.56983 & +28.43353 & $45775_{-3242}^{+1379}$ & $1442.1_{-75.5}^{+42.3}$ & $1.02_{-0.02}^{+0.03}$ & $4.3 \pm 0.1$ & $-1.3 \pm 0.3$ \\
NGC 5824 & 1464 & bPAGB & 225.99601 & $-$33.05991 & *$16822_{-602}^{+630}$* & *$3269.3_{-261.9}^{+273.9}$* & *$6.78 \pm 0.24$* & *$4.3_{-0.6}^{+0.5}$* & *$-1.9 \pm 0.3$* \\
NGC 5986 & 6225 & yPAGB & 236.52059 & $-$37.78401 & $6538_{-101}^{+94}$ & $1688.5_{-36.5}^{+36.4}$ & $32.06_{-1.05}^{+1.16}$ & $1.6 \pm 0.2$ & $-1.3 \pm 0.3$ \\
NGC 5986 & 7430 & yPAGB & 236.51398 & $-$37.77893 & $8229_{-49}^{+64}$ & $1918.6_{-37.2}^{+39.9}$ & $21.58_{-0.42}^{+0.38}$ & $2.8 \pm 0.2$ & $-1.5 \pm 0.3$ \\
NGC 6093 (M80) & 4862 & bPAGB & 244.26753 & $-$22.98895 & $14963_{-261}^{+255}$ & $904.8_{-36.8}^{+30.4}$ & $4.49_{-0.10}^{+0.11}$ & $4.0_{-0.9}^{+0.7}$ & $-1.8 \pm 0.3$ \\
NGC 6205 (M13) & 17229 & vPAGB & 250.39033 & +36.43558 & $24251_{-574}^{+582}$ & $2498.1_{-170.3}^{+129.2}$ & $2.92 \pm 0.08$ & $4.3_{-0.8}^{+0.5}$ & $-1.6 \pm 0.3$ \\
NGC 6273 (M19) & 18767 & bPAGB & 255.66320 & $-$26.25817 & $13243_{-230}^{+240}$ & $2072.3_{-76.0}^{+86.0}$ & $8.71 \pm 0.21$ & $2.6_{-0.5}^{+0.8}$ & $-1.7 \pm 0.3$ \\
NGC 6273 (M19) & 24885 & yPAGB & 255.64666 & $-$26.25672 & $6257_{-101}^{+107}$ & $1734.6_{-43.5}^{+44.6}$ & $35.46_{-1.45}^{+1.49}$ & $1.7 \pm 0.2$ & $-1.5 \pm 0.3$ \\
NGC 6402 (M14) & 7228 & bPAGB & 264.38825 & $-$3.24770 & $15341_{-213}^{+227}$ & $1822.0_{-44.1}^{+44.4}$ & $6.09 \pm 0.14$ & $4.4_{-0.6}^{+0.4}$ & $-1.3 \pm 0.3$ \\
NGC 6712 & 49 & bPAGB & 283.27414 & $-$8.71054 & $11084_{-288}^{+542}$ & $1321.5_{-76.6}^{+132.1}$ & $10.04 \pm 0.21$ & $1.8_{-0.2}^{+0.4}$ & $-0.9 \pm 0.3$ \\
NGC 6779 (M56) & 338 & vPAGB & 289.17201 & +30.21347 & $28042_{-674}^{+688}$ & $1983.5_{-145.2}^{+107.6}$ & $1.97 \pm 0.06$ & $4.2_{-0.6}^{+0.5}$ & $-2.0 \pm 0.3$ \\
\enddata
\tablecomments{Posterior median effective temperatures, bolometric luminosities, radii, surface gravities, and metal abundances, with 68\% credible intervals.\\
$*$ Indicates a post-AGB star with tentative fit parameters that do not have UV photometric coverage (\textit{Swift} or GALEX). These will be corrected when data becomes available.\\
$^a$ Star ID is the listed value from the \textit{uBVI} catalog.\\
$^b$ PAGB classification described in Section~\ref{sec:PAGBclass}: yellow post-AGB (yPAGB), blue post-AGB (bPAGB), violet post-AGB (vPAGB).}
\end{deluxetable*}

\subsection{Limitations of the Analysis}\label{sec:limits}
One significant challenge in modeling SEDs is appropriately accounting for dust extinction. This is especially important in the UV where extinction effects are strongest and small uncertainties in $E(B-V)$ can lead to large errors in the derived temperatures and luminosities for UV-bright stars. There are several extinction laws that depend on dust composition, density, and grain size, with variations seen between galaxies and within our own Galaxy \citep{Peek2013, Gordon2023}. While this study adopted the \citetalias{Cardelli1989} extinction law with a Milky Way diffuse $R_V = 3.1$, we find that clusters with higher $E(B-V)$ show systematically poorer SED fits in the UV\null. This behavior suggests that alternative extinction curves or non-standard values of $R_V$ may be necessary (see Appendix~\ref{app:extinction}).

Additionally, several GCs in our sample exhibit evidence for reddening gradients across the face of the cluster \citep{Pancino2024}. Currently, we do not account for this in our SED fitting. The level of differential reddening, listed under the $dE(B-V)$ column in Table~\ref{tab:GCParameters}, varies but can be considerable for some systems (e.g., $dE(B-V)=0.266$ in NGC~6712). Incorporating spatially resolved reddening corrections or maps will be important for improving parameter estimates, particularly for objects with a greater flux distribution in the UV.

Another limitation is the assumptions of local-thermodynamic equilibrium (LTE) and plane-parallel atmospheres, which are used in the ATLAS9 models.  The results from \citet{Sahin2009} (discussed in Section~\ref{sec:M79pagb}) indicate that PAGB stars may violate these assumptions. Non-LTE corrections could be necessary, especially for the hotter stars where these effects are more significant (see, e.g., Section~\ref{sec:NGC6712pagb}).

Finally, to be a primary distance indicator, PAGB stars need precise parallaxes to calibrate their intrinsic luminosities. However, as discussed in Section~\ref{sec:membership}, \textit{Gaia} parallaxes are less reliable in dense stellar clusters due to crowding and photometric blending. In NGC~6273, for example, the two PAGB stars have significantly different parallax measurements --- $0.1348 \pm 0.0157$ mas and $0.2027 \pm 0.0236$ mas --- despite belonging to the same cluster. The latter value is inconsistent with the cluster distance of 8.527 kpc ($\approx 0.1173\pm 0.0028$ mas). Given these limitations, we assumed the published literature distances listed in Table~\ref{tab:GCParameters} when deriving stellar radii from the model scale factor and for computing luminosities. Future astrometric datasets, such as later \textit{Gaia} data releases, may improve on individual parallax measurements to be used directly in the SED analysis.

\section{PAGB Luminosity Function}\label{sec:Lfunction} 
\begin{figure*}[htbp]
    \centering
    \includegraphics[width=\linewidth]{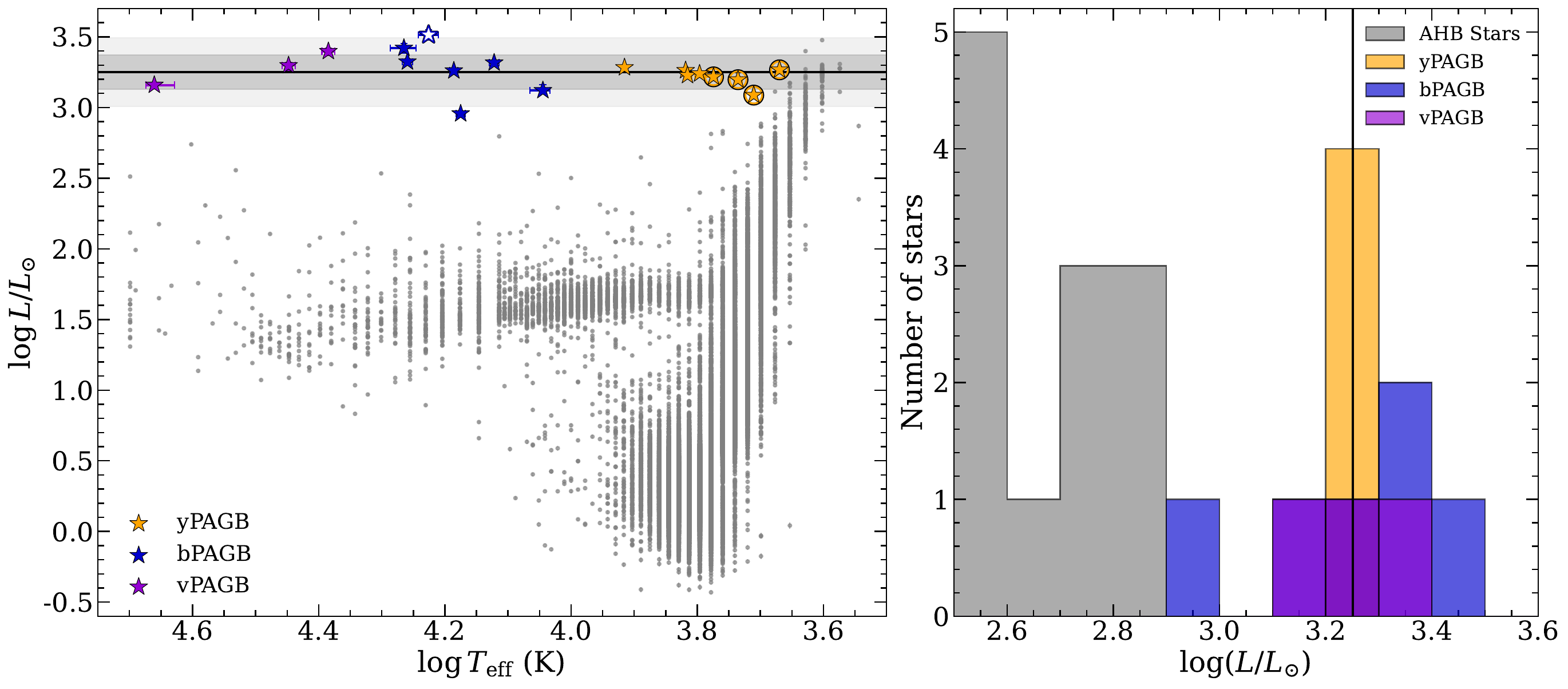}
    \caption{Left: HR Diagram for yPAGB, bPAGB, and vPAGB stars (yellow, blue, and violet star markers, respectively).  The gray circles show the best-fit positions for other reliable ($\chi^2_{\nu} < 10$) members of the PAGB-hosting clusters; these were fit using only the $\chi^2$ grid search and are not modeled with the Bayesian SED-fitting procedure. The vertical striations in their temperatures is a result of the finite model grid resolution, which was not interpolated for these fits. Most uncertainties are smaller than the plot markers. PAGB stars with unfilled markers indicate incomplete UV photometric coverage and will have updated SED fits when additional data becomes available. Known variable stars are enclosed with circle symbols and are excluded from the measurement of the luminosity function. The yPAGB stars occupy cooler temperatures, $3.70 < \log{T_{\mathrm{eff}}} < 3.95$, while bPAGB and vPAGB stars define the hotter end of the distribution, $4.05 < \log{T_{\mathrm{eff}}} < 4.66$. All PAGB stars occupy a narrow luminosity range centered at $\log{(L_{\mathrm{bol}}/L_{\odot})} = 3.25$ with a dispersion of 0.12 dex. Shaded regions indicate $\pm 1\sigma$ and $\pm 2\sigma$ deviations around the mean. Right: Observed luminosity function of the non-variable PAGB sample used in this analysis. Bright, $\log{(L_{\mathrm{bol}}/L_{\odot})} > 2.5$, AHB members are shown in gray.}
    \label{fig:HR}
\end{figure*}

Figure~\ref{fig:HR} shows the locations of the PAGB stars in the HR diagram, along with other cluster members for those systems containing PAGB candidates.  Only objects with SED fits that satisfy $\chi^2_{\nu} < 10$ have been plotted.  Clearly, the GC PAGB stars occupy a very narrow range of luminosity range, with $\langle \log{(L_{\mathrm{bol}}/L_{\odot})} \rangle = 3.25$ and a dispersion of 0.12 dex. Known variable stars (circled) are shown but excluded from the measurement of the luminosity function, as their flux variations and the amplitude of their pulsations are larger than the expected intrinsic scatter. Moreover, the SED fitting is based on photometry taken at different epochs and sample different pulsation phases, introducing additional uncertainty in the fitted parameters.

The yPAGB stars are even more tightly grouped, at lower temperature range ($3.70 < \log{T_{\mathrm{eff}}} < 4.00$) and $\log{(L_{\mathrm{bol}}/L_{\odot})} = 3.25 \pm 0.02$. The bPAGB and vPAGB stars extend over $4.00 < \log{T_{\mathrm{eff}}} < 4.66$. Because these stars are so hot, their SEDs rise steeply in the UV, making near and far-UV photometry critical for accurately constraining the objects' temperatures and bolometric luminosities. One PAGB candidate (shown with an unfilled symbol in Figure~\ref{fig:HR}) is excluded from the calculation of the mean luminosity and its dispersion because it lacks UV photometry. Additional observations will enable a more robust constraint of its stellar properties and it will be included in the sample in future work.

\begin{figure*}[htbp]
    \centering
    \includegraphics[width=\linewidth]{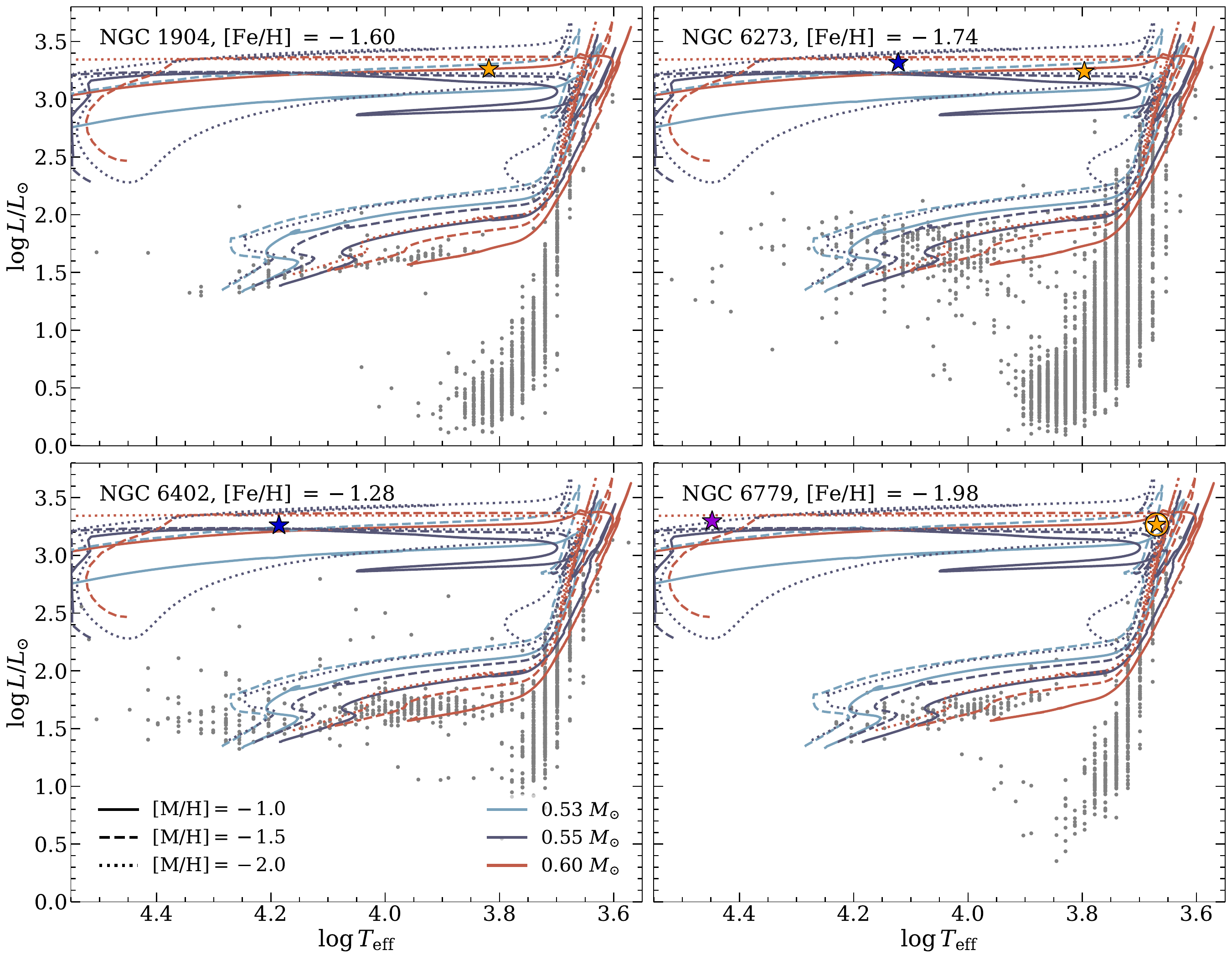}
    \caption{HR diagrams of the PAGB stars in NGC~1904 (top-left), NGC~6273 (top-right), NGC~6402 (bottom-left), and NGC~6779 (bottom-right) alongside theoretical evolutionary tracks from \citetalias{Moehler2019}. Models begin at the ZAHB and include three metallicities ($\mathrm{[Fe/H]} = -1.0, -1.5, -2.0$; solid, dashed, and dotted lines) and three ZAHB stellar masses ($0.53$, $0.55$, and $0.60~M_{\odot}$; distinguished by color). All PAGB stars lie in the region covered by several post-HB models, and are consistent with predicted PAGB luminosities and temperatures of low-mass stars in metal-poor systems.}
\label{fig:PAGBtracks}
\end{figure*}

Figure~\ref{fig:PAGBtracks} compares the observed PAGB stars in NGC~1904, NGC~6273, NGC~6402, and NGC~6779 with post-HB evolutionary tracks from \citetalias{Moehler2019}. All the PAGB stars are consistent with predicted low-mass post-HB models and theoretical expectations for PAGB stars. Multiple tracks overlap at these luminosities and temperatures, indicating that more than one combination of mass and metal abundance can reproduce their observed positions on the HR diagram.

Based on the dependence of PAGB evolutionary pathways and the timescales associated with HB morphology (discussed in Section~\ref{sec:postHBevolution}), AHB and PAGB stars are expected to occur preferentially in metal-poor systems with substantial BHB populations. Consistent with this expectation and the observational result of \citetalias{Davis2022}, 14 of the 16 PAGB-hosting clusters (including rPAGB and variable PAGB) have intermediate to low metallicities ($\mathrm{[Fe/H]} \lesssim -1.28$), and 12 of the 14 clusters with measured HB ratios have predominantly blue HBs ($\mathrm{HBR} > 0$) --- reinforcing the dependence between HB morphology, metallicity, and the frequency of PAGB stars. The PAGB luminosities show no clear dependence on cluster metallicity across our sample. However, this conclusion is limited to the small number of higher-metallicity clusters found to host PAGB stars.

\section{Discussion}\label{sec:discussion}
Early surveys of stars in Galactic GCs laid the foundation for identifying PAGB candidates. The UV-bright star survey of \citet{Zinn1972} (hereafter ZNG72) cataloged 156 bright, non-variable candidates using Johnson $U$-band observations of 27 GCs. Many of these were later confirmed to be evolved post-HB objects. A subsequent spectroscopic study by \citet{deBoer1985} used the International Ultraviolet Explorer \citep[IUE;][]{Boggess1978} observations to determine the effective temperatures, gravities, and luminosities of 10 PAGB candidates in seven clusters. 

More recent work has expanded the census of PAGB stars and characterized their properties. \citetalias{Moehler2019} studied previously known and newly identified PAGB stars using wide-field UV imaging and spectroscopy obtained with the FOcal Reducer and low-dispersion Spectrograph 2 \citep[FORS2;][]{Appenzeller1998} at the ESO Very Large Telescope (VLT\null).  \citetalias{Davis2022} conducted the first cluster-wide photometric search for intermediate-temperature PAGB stars across 97 Galactic GCs, identifying nine yPAGB and four rPAGB stars across ten GCs.

\begin{figure*}[htbp]
    \gridline{
    \fig{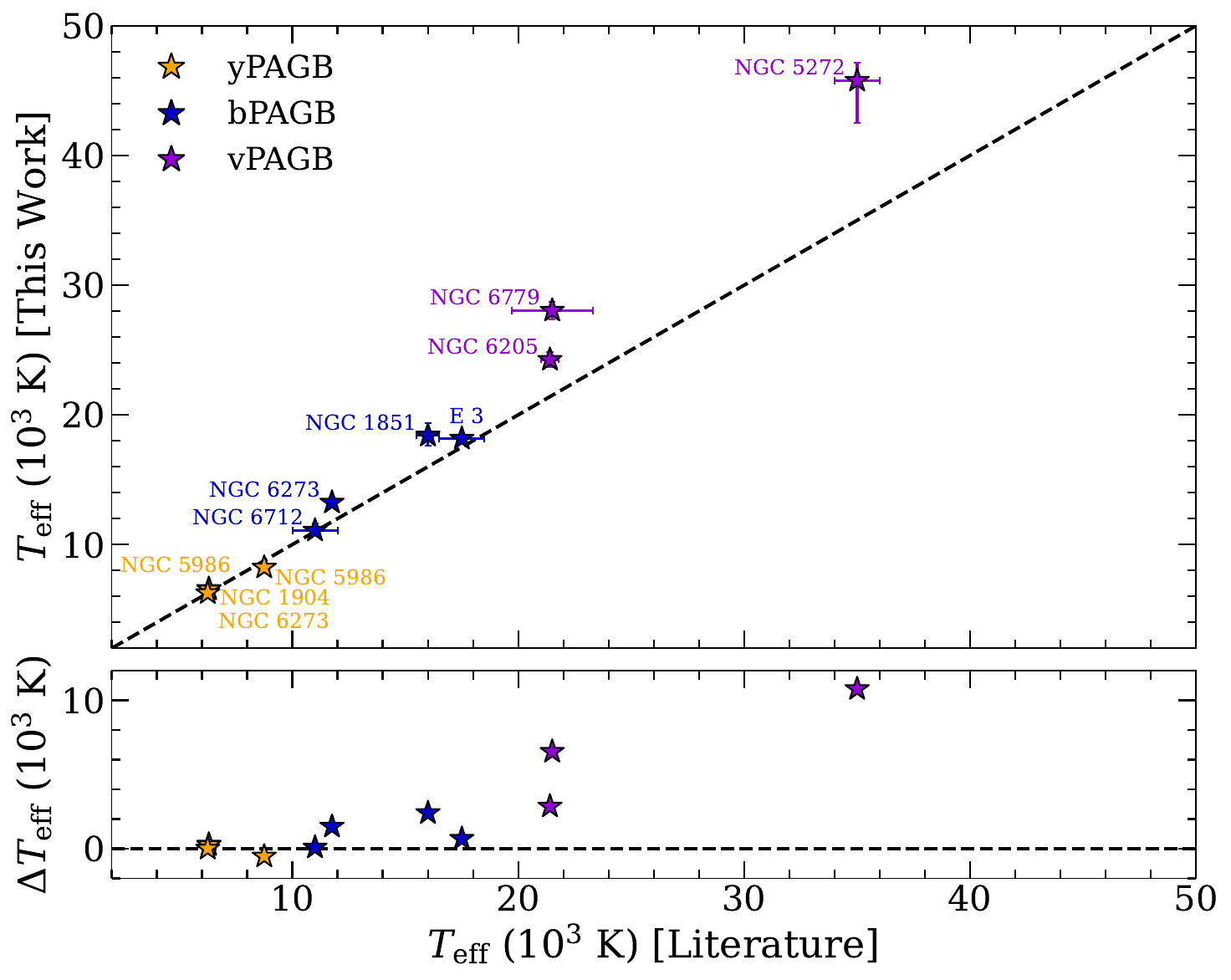}{0.48\textwidth}{(a)}
    \fig{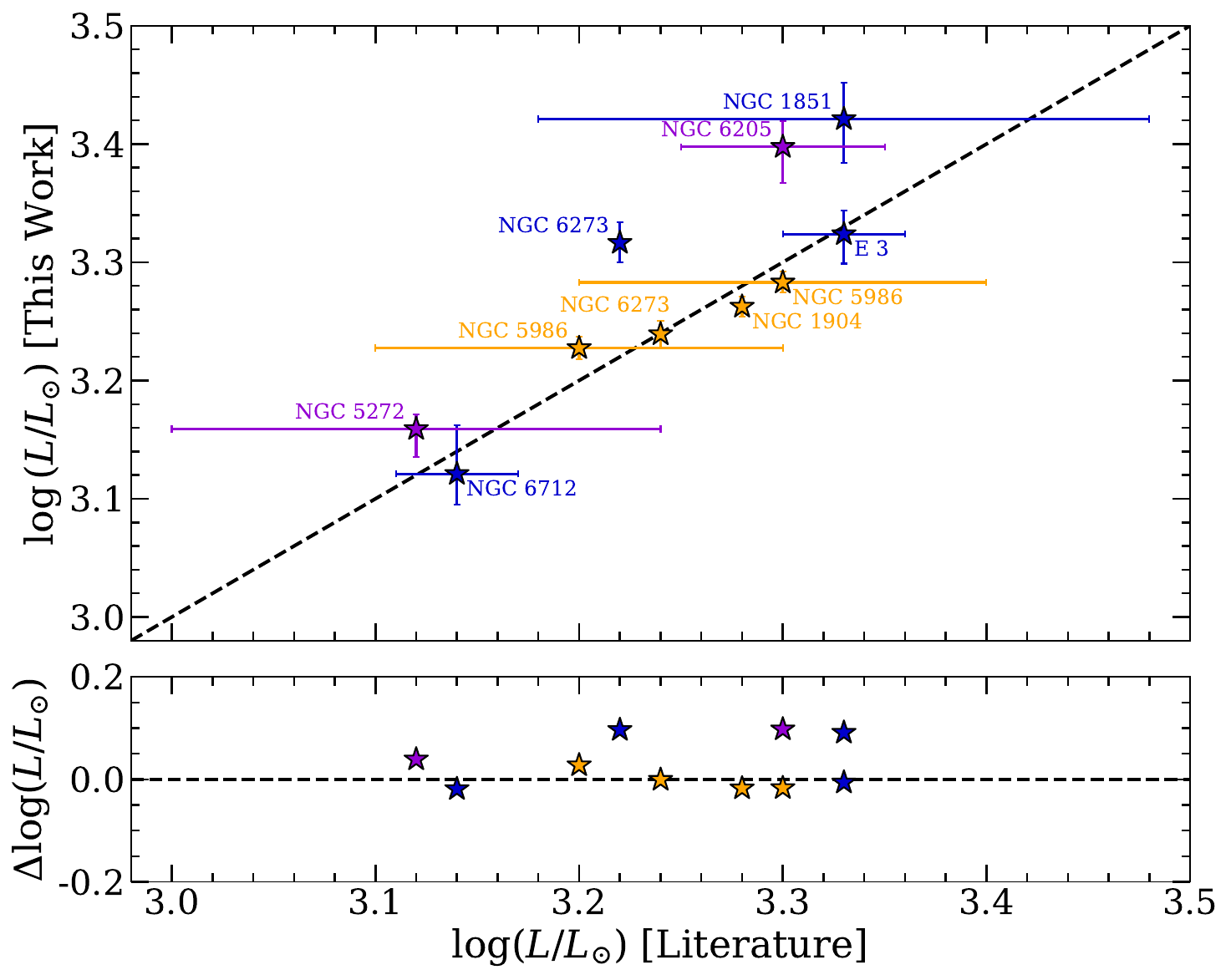}{0.48\textwidth}{(b)}
    }
    \caption{Comparison between literature values (including both spectroscopic and photometric studies, when available) and the results from this work for the inferred effective temperature (left) and bolometric luminosity (right) of the PAGB stars in this sample. Most objects are consistent with previously published values within the quoted uncertainties. The largest discrepancies occur for the hot vPAGB stars in NGC~5272, NGC~6205, and NGC~6779, for which we infer higher effective temperatures, and for the bPAGB and vPAGB stars in NGC~6205 and NGC~6273, which we derive slightly higher luminosities. These differences likely reflect the limitations of broadband SED fitting for very hot objects with steep UV slopes.}
    \label{fig:PAGB_T-L_comp}
\end{figure*}

Building on these efforts, this work provides an analysis of intermediate-temperature and hot PAGB stars by performing UV through IR broadband SED fitting, and applying a consistent methodology to derive temperatures, luminosities, and radii for each object. The following subsections discuss individual PAGB stars in the context of previous work, highlighting agreements and discrepancies. Figure~\ref{fig:PAGB_T-L_comp} presents a visual comparison between the literature values, when available, and results obtained in this work for effective temperature and bolometric luminosity. Overall, our inferred stellar parameters are generally consistent with photometric and spectroscopic studies within the quoted uncertainties, although we derive higher effective temperatures for the hottest PAGB stars whose SEDs peak in the far-UV where our photometric coverage is limited.

\subsection{bPAGB Star in E3 (ESO 37-1)} % T = 18171, logL = 3.32, log g = 3.9
The bPAGB star in the sparse, relatively metal-rich, $\mathrm{[Fe/H]} = -0.83$, cluster E3 was first identified by \citet{Kumar2024}. They performed SED fitting using photometry spanning UV to IR wavelengths, and derived $T_{\mathrm{eff}} = 17{,}500 \pm 1000$ K, $\log{g} = 2.37 \pm 0.2$, and $\log{(L/L_{\odot})} = 3.33 \pm 0.03$. Our results are in excellent agreement with their temperature and luminosity estimates, and we infer a higher surface gravity. As discussed in Section~\ref{sec:M-logg}, broadband photometric measurements of hot stars provide only a weaker constraint on $\log{g}$.

\subsection{bPAGB Star in NGC~1851} % T = 18407, logL = 3.42, log g = 2.4
The bPAGB star in NGC~1851 was first designated as UV5 in \citet{Vidal1975}, and later analyzed photometrically and spectroscopically by \citet{Parise1994A} and \citet{Dixon1994}. \citet{Parise1994A} found the object to be consistent with an evolved PAGB star, and \citet{Dixon1994} derived $T_{\mathrm{eff}} = 16{,}000 \pm 500$ K, $\log{(L/L_{\odot})} = 3.33 \pm 0.15$, and $\log{g} = 2.5 \pm 0.25$ using UV spectra taken with the Hopkins Ultraviolet Telescope \citep[HUT;][]{Davidsen1992}. Our results are consistent with the published luminosity and surface gravity estimates, and we derive a moderately higher effective temperature.

\subsection{yPAGB Star in NGC~1904 (M79)}\label{sec:M79pagb} % T = 6584, logL = 3.26, log g = 1.3
The yPAGB star in NGC~1904 was first detected by \citet{Bond2016}, who fit a $T_{\mathrm{eff}}=6300$ K blackbody spectrum to UV through IR photometry, including 2MASS and the Wide-field Infrared Survey Explorer \citep[WISE;][]{Wright2010} observations. A high-resolution spectroscopic analysis by \citet{Sahin2009} found a consensus model of $T_{\mathrm{eff}}=6300$ K and $\log{g} = 0.8$, consistent with an evolved low-surface gravity object, but with an $\mathrm{[Fe/H]} = -2.0$ that is $\sim 0.5$ dex lower than that of the cluster's red giant population. The authors also showed that, under the assumption of LTE, higher temperatures and surface gravities were needed to match the PAGB star's metallicity to that of the cluster RGB stars. They therefore concluded that the inconsistency was likely due to limitations in classical stellar atmosphere models, such as the assumption of LTE, or a temperature gradient that deviates from standard models. 

Our results are consistent with the published temperature and luminosity estimates within the uncertainties. We also derive a similar low surface gravity, although the spectroscopic value from \citet{Sahin2009} is likely more reliable.

\subsection{vPAGB Star in NGC~5272 (M3)} % T = 45775, logL = 3.16, log g = 4.3
The vPAGB star in NGC~5272 is the brightest cluster member, was first identified by \citet{VonZeipel1908} (vZ 1128) and later designated ZNG 1 in \citetalias{Zinn1972}. UV spectra obtained by \citet{Dixon1994} with the HUT yielded $T_{\mathrm{eff}} = 35{,}000 \pm 1000$ K, $\log{(L/L_{\odot})} = 3.21 \pm 0.12$, and $\log{g} = 4.0 \pm 0.25$ estimates. \citet{Chayer2015} subsequently analyzed
Keck spectra obtained with the High Resolution Echelle Spectrometer (HIRES), finding $T_{\mathrm{eff}} = 36{,}600 \pm 400$ K and $\log{g} = 3.95 \pm 0.10$. We derive a luminosity consistent with the analysis of \citet{Dixon1994}, but infer a higher effective temperature than both previous studies. For vPAGB stars such as this object, the peak of the SED lies at far-UV wavelengths beyond the coverage of our broadband photometry. The spectroscopic temperatures are therefore likely to be more accurate.

\subsection{bPAGB Star in NGC~5824} % T = 16822, logL = 3.51, log g = 4.3
The bPAGB star in NGC~5824 was first identified in the \citetalias{Davis2022} study. To our knowledge, no detailed photometric or spectroscopic analysis has been published for this object. This work provides the first estimates of its stellar parameters.

\subsection{yPAGB Stars in NGC~5904 (M5)} % T = 5134, logL = 3.09, log g = 4.3; T = 5946, logL = 3.22, log g = 4.2
The yPAGB stars in NGC~5904 were identified as variable stars by \citet{Packer1890} and later designated V42 and V84 (IDs 10926 and 6774 in Table~\ref{tab:PAGBcatalog}, respectively). V42 is classified as a W Virginis star with a period $P \approx 25.68$ days, and V84 is an RV Tauri star with $P \approx 26.46$ days \citep{CruzReyes2025}. \citet{Gonzalez1997} obtained spectra of both from the McDonald Observatory 2.1-m telescope. They derived $T_{\mathrm{eff}} = 5750\,\text{--}\,6250$ K and $\log{g} = 0.5\,\text{--}\,1.5$ for V42, and  $T_{\mathrm{eff}} = 5250\,\text{--}\,5400$ K and $\log{g} = 0.3\,\text{--}\,1.1$ for V84. \citet{Carney1998} obtained six spectra for V42 from the 4m telescopes at Cerro Tololo Inter-American Observatory (CTIO) and Kitt Peak National Observatory (KPNO) and the 2.7 m telescope at McDonald Observatory at three different phases of its pulsation cycle and derived $T_{\mathrm{eff}} = 5200\,\text{--}\,6000$ K and $\log{g} = 0.8\,\text{--}\,1.5$. 

Our derived temperatures (not included in Table~\ref{tab:PAGBparams}) are broadly consistent with the range of spectroscopic values, but the fitted surface gravities are substantially higher. However, the broadband photometry used to construct the SEDs was obtained at different epochs and sample different phases of the pulsation cycles, which limits the reliability of the fitted atmospheric parameters for these and other variable stars in our sample.

\subsection{yPAGB Stars in NGC~5986} % T = 8229, logL = 3.28, log g = 2.8; T = 6538, logL = 3.23, log g = 1.6
Two yPAGB stars in NGC~5986 were identified by \citet{Alves2001} as ID 6 and ID 7 (IDs 7430 and 6225, respectively, in this work). \citet{Jasniewicz2004} used high resolution spectra from the UV-Visual Echelle Spectrograph \citep[UVES;][]{Dekker2000} to derive $T_{\mathrm{eff}} = 8750 \pm 250$ K, $\log{g} = 2.0 \pm 0.3$, $\log{(L/L_{\odot})} = 3.3 \pm 0.1$ for ID 6, and $T_{\mathrm{eff}} = 6300 \pm 250$ K, $\log{g} = 1.0 \pm 0.3$, $\log{(L/L_{\odot})} = 3.2 \pm 0.1$ for ID 7. Our results are in good agreement with these published temperatures and luminosities.

\subsection{bPAGB Star in NGC~6093 (M80)} % T = 14964, logL = 2.96, log g = 4.0
The bPAGB star in NGC~6093 was designated ZNG 3 by \citetalias{Zinn1972}. To our knowledge, no detailed photometric or spectroscopic analysis has been published for this object. This work provides the first estimates of its stellar parameters.

\subsection{vPAGB Star in NGC~6205 (M13)} % T = 24250, logL = 3.40, log g = 4.3
The bPAGB star in NGC~6205 was first noted by E. E. Barnard as ``the most striking example'' of a bright, blue star which outshone other cluster members \citep{Barnard1900, Barnard1909, Barnard1914}. It was designated ZNG 1 by \citetalias{Zinn1972}, and multiple spectroscopic studies have established it as a PAGB star with $T_{\mathrm{eff}} \approx 20{,}000 \pm 1000$ K and $\log{g} = 2.95 \pm 0.1$ \citep{Conlon1994, Dixon1998, Moehler1998, Thompson2007}. More recently, \citet{Dixon2019} used archival spectra to infer $T_{\mathrm{eff}} = 21{,}400 \pm 400$ K, $\log{(L/L_{\odot})} = 3.26\,\text{--}\,3.35$, and $\log{g} = 3.10 \pm 0.03$. Our results are broadly consistent with these analyses, although we infer a moderately hotter effective temperature and higher surface gravity.

\subsection{bPAGB and yPAGB Stars in NGC~6273 (M19)} % T = 6257, logL = 3.24, log g = 1.7; T = 13243, logL = 3.32, log g = 2.6
The two PAGB stars in NGC~6273 were first cataloged in \citetalias{Zinn1972} and are designated as ZNG 2 (bPAGB) and ZNG 4 (yPAGB\null). \citet{Bond2021a} modeled their SEDs and derived $\log{(L/L_{\odot})} = 3.22$ and $T_{\mathrm{eff}} = 11{,}750$ K for ZNG 2, and $\log{(L/L_{\odot})} = 3.24$ and $T_{\mathrm{eff}} = 6250$ K for ZNG 4. Their analysis included Hubble Space Telescope (HST) photometry and additional IR coverage with 2MASS and the Spitzer Space Telescope \citep{Werner2004}. For the yPAGB star, their fit included only the UVOT \textit{uvm2} band, noting that the \textit{uvw1} and \textit{uvw2} filters suffer from read-leak contamination that can greatly affect the photometry of cooler stars. This issue is explored in Appendix~\ref{app:UVOTfilters}. 

Within the uncertainties, our results are in excellent agreement with the results for the yPAGB star.  For the hotter bPAGB object, we infer a higher temperature and luminosity, likely reflecting difference in the UV and IR photometric coverage.

\subsection{bPAGB Star in NGC~6402 (M14)} % T = 15341, logL = 2.93, log g = 4.4
The bPAGB star in NGC~6402 was first cataloged by \citetalias{Zinn1972} (designated ZNG 1) and later classified as a hot PAGB star by \citet{Schiavon2012} (identified as NGC~6402-31). To our knowledge, no detailed photometric or spectroscopic analysis has been published for this object. This work provides the first estimates of its stellar parameters.

\subsection{bPAGB Star in NGC~6712}\label{sec:NGC6712pagb} % T = 11084, logL = 3.12, log g = 1.8
The bPAGB star in NGC~6712 was designated ZNG 1 in \citetalias{Zinn1972}. Multiple spectroscopic studies have established it as a low mass PAGB star with $T_{\mathrm{eff}} \approx 11{,}000$ K, $\log{g} \approx 2.1\,\text{--}\,2.5$, and $\log{(L/L_{\odot})} \approx 3.11\,\text{--}\,3.16$ \citep{Mooney2004, Jasniewicz2004}. However, \citetalias{Moehler2019} --- using non-LTE, line-blanketed model atmospheres --- estimated a much higher temperature and surface gravity ($T_{\mathrm{eff}} = 53{,}300 \pm 4000$ K, $\log{g} = 4.5 \pm 0.12$). Our results are consistent with the lower-temperature solutions within the uncertainties.

\subsection{yPAGB and vPAGB Stars in NGC~6779 (M56)} % T = 28042, logL = 3.30, log g = 4.2; T = 4674, logL = 3.27, log g = 4.9
The variable yPAGB star was designated V6 by \citet{Sawyer1940}, classified as an RV Tauri by \citep{Sawyer1949}, and has a period of $P \approx 44.96$ days \citep{CruzReyes2025}. To our knowledge, no detailed photometric or spectroscopic analysis has derived stellar properties for this object. The vPAGB star in NGC~6779 was designated ZNG 2 in \citetalias{Zinn1972}. \citetalias{Moehler2019} estimated $T_{\mathrm{eff}} = 21{,}500 \pm 1800$ K, $\log{g} = 3.07 \pm 0.16$ for this object. We derive a moderately higher effective temperature and surface gravity.

\subsection{yPAGB Star in NGC~7089 (M2)} % T = 5438, logL = 3.20, log g = 1.8
The variable yPAGB star was discovered by amateur astronomer \citet{Chevremont1898} and designated V11 by \citet{Sawyer1935}. It is an RV Tauri star with a period of $P \approx 33.40$ days \citep{CruzReyes2025}. \citet{Gonzalez1997} obtained spectra at two observing epochs from the McDonald Observatory 2.1-m and CTIO 4-m telescopes. They derived $T_{\mathrm{eff}} = 5250\,\text{--}\,5550$ K and $\log{g} = -0.3\,\text{--}\,0.2$. 

Our temperature estimate, based on broadband photometry obtained at different epochs, is consistent with the range of spectroscopic values, but the fitted surface gravity is substantially higher.

\section{Summary \& Future Work}\label{sec:future}
This study establishes the framework and analysis required for a complete census of PAGB stars. Comparisons with previous spectroscopic analyses (Section~\ref{sec:discussion}) demonstrate that our broadband photometric SED fitting can reliably derive effective temperatures and bolometric luminosities across a wide range of PAGB stars using archival and ground-based imaging --- requiring a fraction of the telescope time needed for detailed high-resolution spectroscopic characterization. 

Using 39 Galactic GCs we have:
\begin{enumerate}\setlength\itemsep{1.5pt}
\renewcommand{\labelenumi}{\theenumi)}
    \item built a multi-wavelength catalog containing far-UV through near-IR photometry;
    \item assessed cluster membership probabilities using \textit{Gaia} astrometry to minimize field star contamination;
    \item defined quantitative PAGB classification boundaries and identified 21 candidate PAGB stars in 16 GCs;
    \item fit our broadband photometry to ATLAS9 stellar atmosphere models to infer the stars' temperatures and luminosities;
    \item evaluated degeneracies in estimated stellar metal abundances and surface gravities, demonstrating that broadband photometry alone usually does not uniquely constrain these parameters;
    \item presented this sample's PAGB luminosity function, which has a mean bolometric luminosity of $\log(L_{\mathrm{bol}}/L_{\odot}) = 3.25$ with an observed dispersion of $\sigma = 0.12$ dex;
    \item examined the correlation between PAGB frequency and HB morphology, finding that the majority of PAGB-hosting GCs have $\mathrm{HBR} > 0$ --- consistent with previous expectations;
    \item compared our results with existing literature and showed that the temperatures and luminosities derived from broadband SED fitting are in good agreement with published spectroscopic analyses.
\end{enumerate}
The next step is to apply this methodology to the full Galactic GC catalog, evaluating previously identified PAGB stars and searching for additional candidates. Expanding the sample size, particularly toward the hot-end of the PAGB sequence, is crucial for improving constraints on stellar evolution models and testing the utility of PAGB stars for extragalactic distance measurements. 

The current sample of 13 non-variable, intermediate-temperature and hot PAGB stars with UV--IR photometry is sufficient to constrain the mean bolometric luminosity of the class to $\log(L_{\mathrm{bol}}/L_{\odot}) = 3.25$ ($M_{\mathrm{bol}} = -3.39$), with an observed dispersion of 0.12 dex (0.30 mag) and a standard error of the mean of $0.03$ dex (0.08 mag). The accuracy of these values is limited by the small number of confirmed objects and the temperature range sampled. A more complete survey will robustly constrain the intrinsic scatter of the luminosity function and test dependencies on metallicity or HB morphology.

Finally, a larger and more uniform sample will enable measurements of the frequency of PAGB stars across different cluster environments, the relative populations of vPAGB/bPAGB/yPAGB/rPAGB subclasses, and the evolutionary pathways and timescales of these stars. Ultimately, expanding the sample of PAGB stars is needed to establish their utility as precise extragalactic standard candles.

\subsection{Improved Parameter Estimation and Modeling}
As demonstrated by the analysis of the yPAGB in NGC~1904 by \citet{Sahin2009} and the bPAGB star in NGC~6712 by \citetalias{Moehler2019}, standard models like ATLAS9 may not fully capture the atmospheric structure and non-LTE effects of these high-luminosity, low-gravity stars.  More sophisticated models may be necessary to reliably determine the stellar properties for stars where these deviations are significant. High-resolution spectroscopic follow-up of PAGB candidates will provide the best constraints on $T_{\mathrm{eff}}$, $\log{g}$, and [Fe/H], and enable direct measurements of line profiles, ionization balance, and elemental abundances. These observations will be essential to construct improved models for evolved PAGB objects.

As discussed in Section~\ref{sec:limits} and explored in Appendix~\ref{app:extinction}, assumptions in the reddening parameter $R_V$ can significantly impact SED fitting, particularly at UV wavelengths where the extinction curve is steepest. In this study, we adopt the canonical average Milky Way extinction curve ($R_V = 3.1$). However, deviations from this frequently used standard value can introduce systematic errors in the derived stellar properties. 

\citet{Zhang2023} built a 2D $R_V$ map using color excess ratios of three million stars from Large Sky Area Multi-Object Fiber Spectroscopy Telescope (LAMOST\null). They found an average value of $R_V = 3.25 \pm 0.25$ and demonstrated there was significant spatial variation across the Milky Way. Their results support the need to consider a varying extinction law, rather than adopting a single fixed $R_V$.

Extinction can be measured by comparing the observed and intrinsic spectra of stars whose SEDs are well-characterized. BHB stars, whose temperatures and surface gravities can be reliably determined from optical photometry, offer a promising probe of local reddening and the variation in $R_V$.  In a future work we will use BHB stars in a set of GCs to refine their cluster-specific reddening laws, provide constraints on UV extinction, and improve the SED fits for the PAGB stars.

\begin{acknowledgments}
This work was supported by the NSF through grant AST2206090.  This work was also supported by NASA grants 80NSSC26K0646 and 80NSSC25K7804. The Institute for Gravitation and the Cosmos is supported by the Eberly College of Science and the Office of the Senior Vice President for Research at the Pennsylvania State University.
\end{acknowledgments}

\facilities{\textit{GALEX}, \textit{Swift}, CTIO: 0.9 m and 1.5 m, KPNO: 0.9 m and 4 m, \textit{Gaia}, 2MASS}

\software{astropy \citep{2013A&A...558A..33A,2018AJ....156..123A,2022ApJ...935..167A}, pandas \citep{2022zndo...3509134T}, NumPy \citep{2020Natur.585..357H}, SciPy \citep{2020NatMe..17..261V}, Matplotlib \citep{Hunter:2007}}

\appendix
\section{Dust Extinction}\label{app:extinction}

\begin{figure}[htbp]
    \centering
    \added{\includegraphics[width=0.75\linewidth]{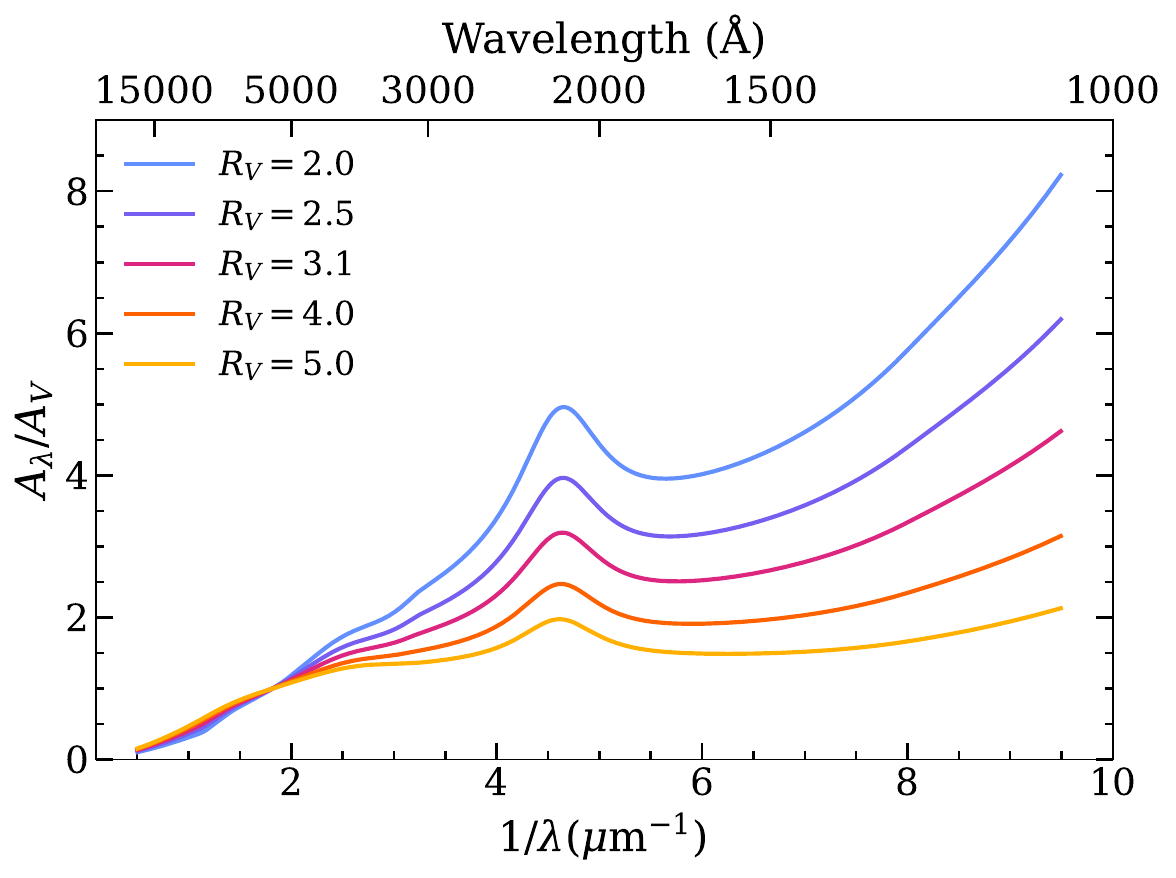}}
    \caption{\citetalias{Cardelli1989} extinction curve for several values of $R_V$ (distinguished by color), normalized such that $A_V = 1$. There is a steep rise in $A_{\lambda}$ toward the UV and a broad 2175 \AA~bump. Increasing $R_V$ produces a flatter (grayer) curve. Overall, the sensitivity to $R_V$ is greatest in the UV and minimal at longer wavelengths.}
    \label{fig:CCM_ext}
\end{figure}

Interstellar extinction varies strongly with wavelength, significantly increasing toward the UV\null. In this work, we applied the parameterized extinction curve of \citetalias{Cardelli1989}, which expresses the total extinction $A_{\lambda}$ as a function of wavelength, and the selective extinction ratio
\begin{equation}
    R_V = \frac{A_V}{E(B-V)}
\end{equation}
for which we choose the canonical $R_V = 3.1$ value. 

Figure~\ref{fig:CCM_ext} illustrates the \citetalias{Cardelli1989} extinction curve for several values of $R_V$; $A_{\lambda}$ is normalized by the $V$-band extinction such that $A_V = 1$. The curve shows several key features: 1) a steep rise in the UV where dust absorption is greatest; 2) a variation with $R_V$ that is greatest in the UV, where increasing $R_V$ produces a flatter (grayer) extinction curve, and the dependence becomes weaker at $\lambda \gtrsim 4000$ \AA; 3) a broad ($\sim 350$~\AA) bump centered on 2175 \AA, most likely a characteristic of dust composition containing graphite or polycyclic aromatic hydrocarbons (PAHs). The latter feature is prominent in the Milky Way, but varies in strength for different environments (e.g., SMC or LMC).

\begin{figure}[htbp]
    \centering
    \includegraphics[width=1.0\linewidth]{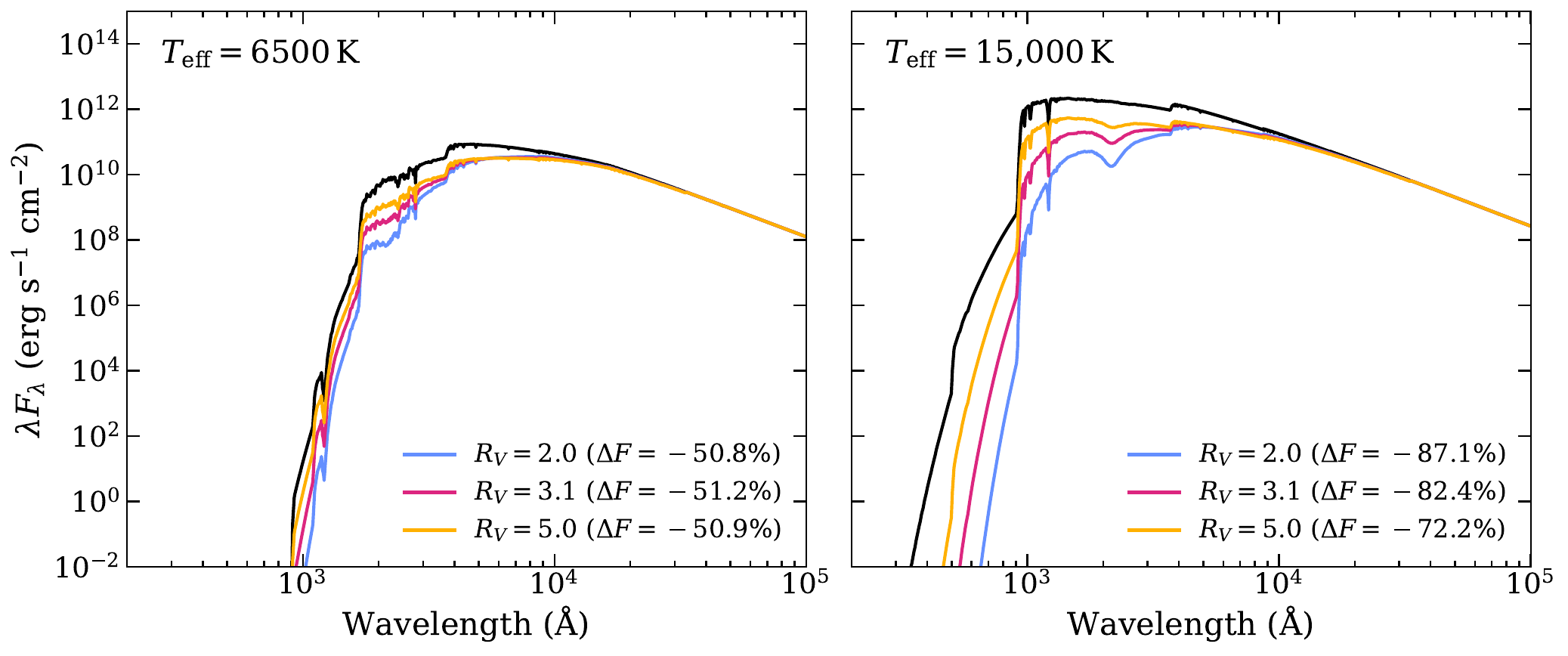}
    \caption{Effect of varying $R_V$ on two model stellar spectra ($T_{\mathrm{eff}} = 6500$ K and $15{,}000$ K) reddened with $A_V = 1$. Changes to the total integrated flux ($\Delta F$) are indicated for each $R_V$. Cooler stars show relatively uniform flux suppression across optical and IR wavelengths, with minor dependence on $R_V$. In contrast, hot stars exhibit substantially greater flux suppression that increases with lower $R_V$. The extinction curve is extended below 1000 \AA, assuming a power law.}
    \label{fig:ext-SED}
\end{figure}
Figure~\ref{fig:ext-SED} demonstrates how the choice of $R_V$ can influence the SED of a cool ($T_{\mathrm{eff}} = 6500$ K) and hot ($T_{\mathrm{eff}} = 15,000$ K) star. Each model is reddened assuming $A_V = 1$ mag using a \citetalias{Cardelli1989} extinction curve, but uses a different values of $R_V$.  For cooler stars, which emit most of their flux in the optical and IR wavelengths, the SED is relatively unaffected by the assumption of an $R_V$: The fraction of integrated flux decreases by $\approx 50\%$, but is nearly constant across all $R_V$. For hot stars, whose fluxes peak in the near-UV, extinction reddens the SED significantly below $\lambda \sim 3000$ \AA\null. Decreasing $R_V$ produces a large suppression of flux that grows toward shorter wavelengths;  nearly $90\%$ of the flux is suppresed for $R_V = 2.0$, and the UV 2175~\AA\ bump is increased strength.

Figures \ref{fig:CCM_ext} and \ref{fig:ext-SED} together show that $R_V$ can significantly affect SED fitting, particularly for hot PAGB (bPAGB and vPAGB) stars. Incorrect assumptions about the extinction law can translate into systematic shifts in temperature and luminosity.  Fortunately, the HB stars of a GC (and, in particular, its blue HB stars) may be useful in constraining the $R_V$ value applicable to the cluster's PAGB stars. This will be explored in future work.

\section{\textit{Swift} UVOT Filter Red-Leak}\label{app:UVOTfilters}
The \textit{Swift} UVOT \textit{uvw2} and \textit{uvw1} filters are known to have a small red-leak that can be particularly significant for cool stars where the optical flux contribution is important. To evaluate the sensitivity of our SED fits to this effect we constructed an ersatz function of the \textit{uvw2} and \textit{uvw1} transmission curves, suppressing the blue response by 20\% and leaving the red end unchanged.

\begin{figure}[htbp]
    \centering
    \includegraphics[width=1.0\linewidth]{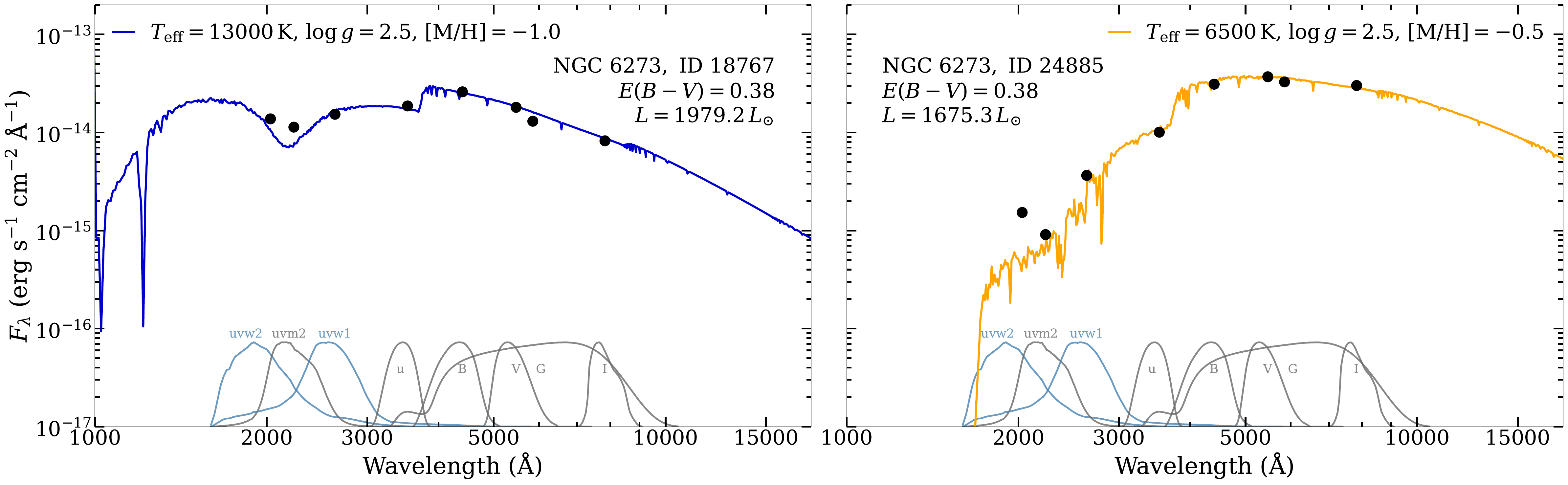}
    \caption{SED fits for bPAGB (left) and yPAGB (right) stars in NGC~6273 using the ersatz \textit{uvw2} and \textit{uvw1} filter curves, in which the blue response is suppressed by 20\% while the red end is unchanged. The unmodified filter transmission curves are shown at the bottom of each panel, with the \textit{uvw2} and \textit{uvw1} curves highlighted in light blue and the remaining shown in gray.}
    \label{fig:SED-ersatz}
\end{figure}

We then refit both the yPAGB and bPAGB stars of NGC~6273 using this modified filter response, keeping all other photometry and procedures the same. The resulting best-fit models are shown in Figure~\ref{fig:SED-ersatz}. The luminosity decreased by only $5.2~L_{\odot}$ (0.3\%) for the yPAGB, and increased by $54.7~L_{\odot}$ (2.8\%) for the bPAGB\null. There is no variation in the fitted $T_\mathrm{eff}$ and $\log{g}$, and the changes in the best-fit $\chi^2_{\nu}$ are negligible.

These tests indicate that the red-leak has a minimal impact on our SED fitting and derived stellar parameters for PAGB stars. Even under an extreme ersatz curve with 20\% blue response suppression, the inferred stellar parameters remain largely unchanged. Any luminosity variations are well below the uncertainties.

\bibliography{ref}{}
\bibliographystyle{aasjournalv7.1}

\end{document}